\documentclass[fleqn,usenatbib]{mnras}

\usepackage{newtxtext,newtxmath}
\usepackage[T1]{fontenc}

\DeclareRobustCommand{\VAN}[3]{#2}
\let\VANthebibliography\thebibliography
\def\thebibliography{\DeclareRobustCommand{\VAN}[3]{##3}\VANthebibliography}

\usepackage{graphicx}	% Including figure files
\usepackage{amsmath}	% Advanced maths commands

\usepackage{orcidlink}

\newcommand{\rev}[1]{\textcolor{black}{#1}}
\newcommand{\revtwo}[1]{\textcolor{black}{#1}}

\title[CEERS Galactic Dwarfs]{Cosmic Evolution Early Release Science Survey (CEERS): Multi-classing Galactic Dwarf Stars in the deep JWST/NIRCam }

\author[B.W. Holwerda]{
B.W. Holwerda$^{1}$\orcidlink{0000-0002-4884-6756}, 
\thanks{E-mail: benne.holwerda@louisville.edu}
Chih-Chun Hsu$^{2}$\orcidlink{0000-0002-5370-7494},
Nimish Hathi$^{3}$\orcidlink{0000-0001-6145-5090}, % (nphathi@gmail.com)
Laura Bisigello$^{4,5}$\orcidlink{0000-0003-0492-4924},  % <laura.bisigello@inaf.it>
Alexander de la Vega$^{6}$, % <alexandd@ucr.edu
\and
Pablo Arrabal Haro$^{7}$\orcidlink{0000-0002-7959-8783}, % (parrabalh@gmail.com)
Micaela Bagley$^{8}$\orcidlink{0000-0002-9921-9218}, % (mbagley@utexas.edu)
Mark Dickinson$^{7}$\orcidlink{0000-0001-5414-5131}, % (mark.dickinson@noirlab.edu)
Steven L. Finkelstein$^{8}$\orcidlink{0000-0001-8519-1130}, % (stevenf@astro.as.utexas.edu)
\and
Jeyhan S. Kartaltepe$^{9}$\orcidlink{0000-0001-9187-3605}, % (jeyhan@astro.rit.edu)
Anton M. Koekemoer$^{3}$\orcidlink{0000-0002-6610-2048}, % (koekemoer@stsci.edu)
Casey Papovich$^{10,11}$\orcidlink{0000-0001-7503-8482}, % (papovich@tamu.edu)
Nor Pirzkal$^{3}$\orcidlink{0000-0003-3382-5941}, % (npirzkal@stsci.edu)
Kyle Cook$^{1}$,
\and 
Clayton Robertson$^{1}$\orcidlink{0000-0002-5404-1372},
Caitlin M Casey$^{8}$\orcidlink{0000-0001-7503-8482},
Christian Aganze$^{12}$\orcidlink{},
Pablo G. P\'erez-Gonz\'alez$^{13}$\orcidlink{0000-0003-4528-5639}, 
\and
Ray A. Lucas$^{3}$\orcidlink{0000-0003-1581-7825},  
Shardha Jogee$^{8}$\orcidlink{0000-0002-1590-0568}, 
Stephen Wilkins$^{14}$\orcidlink{0000-0003-3903-6935}, 
Denis Burgarella$^{15}$,
and
Allison Kirkpatrick$^{16}$\orcidlink{0000-0002-5537-8110}\\
$^{1}$ Department of Physics and Astronomy, University of Louisville, Natural Science Building 102, 40292 KY, Louisville, USA\\
$^{2}$ Center for Interdisciplinary Exploration and Research in Astrophysics (CIERA)\\
$^{3}$ Space Telescope Science Institute (STSCI), 3700 San Martin Dr., Baltimore, 21218 MD, USA\\
$^{4}$ Dipartimento di Fisica e Astronomia "G.Galilei", Universit\'a di Padova, Via Marzolo 8, I-35131 Padova, Italy\\
$^{5}$ INAF--Osservatorio Astronomico di Padova, Vicolo dell'Osservatorio 5, I-35122, Padova, Italy\\
$^{6}$ Department of Physics and Astronomy, University of California, 900 University Ave, Riverside, CA 92521, USA\\
$^{7}$ NSF's National Optical-Infrared Astronomy Research Laboratory, 950 N. Cherry Ave., Tucson, AZ 85719, USA\\
$^{8}$ Department of Astronomy, The University of Texas at Austin, Austin, TX, USA\\
$^{9}$ Laboratory for Multiwavelength Astrophysics, School of Physics and Astronomy, Rochester Institute of Technology,\\ 84 Lomb Memorial Drive, Rochester, NY 14623, USA\\
$^{10}$ Department of Physics and Astronomy, Texas A\&M University, College Station, TX, 77843-4242, USA\\
$^{11}$ George P.\ and Cynthia Woods Mitchell Institute for Fundamental Physics and Astronomy, Texas A\&M University, College Station, TX, 77843-4242 USA\\
$^{12}$ Department of Physics \& Astronomy, University of California, San Diego, CA 92093, USA\\
$^{13}$ Centro de Astrobiolog\'ia (CAB), CSIC-INTA, Ctra. de Ajalvir km 4, Torrej\'on de Ardoz, E-28850, Madrid, Spain\\
$^{14}$ Department of Physics and Astronomy, University of Sussex, Brighton BN1 9QH, United Kingdom \\
$^{15}$Laboratoire d'Astrophysique de Marseille,  Aix-Marseille Universite, CNRS Technopole de Chateau-Gombert 38, rue Frederic Joliot-Curie, \\13013 Marseille, France\\
$^{16}$Department of Physics and Astronomy, University of Kansas, Lawrence, KS 66045, USA\\
}
\date{Accepted XXX. Received YYY; in original form ZZZ}

\pubyear{2022}

\begin{document}
\label{firstpage}
\pagerange{\pageref{firstpage}--\pageref{lastpage}}
\maketitle

% Abstract of the paper
\begin{abstract}

Low mass (sub)stellar objects represent the low end of the initial mass function, the transition to free-floating planets and a prominent interloper population in the search for high-redshift galaxies. Without proper motions or spectroscopy, can one identify these objects photometrically? JWST/NIRCam has several advantages over HST/WFC3 NIR: more filters, a greater wavelength range, and greater spatial resolution.
Here, we present a catalogue of (sub)stellar dwarfs identified in the Cosmic Evolution Early Release Science Survey (CEERS). We identify 518 stellar objects down to mF200W~28 using half-light radius, a full three magnitudes deeper than typical HST/WFC3 images. A kNN nearest neighbour algorithm identifies and types these sources, using four HST/WFC3 and four NIRCam filters, trained on SpeX spectra of nearby brown dwarfs.  
The kNN with four neighbors classifies well within two subtypes: e.g M2$\pm$2 or T4$\pm$2, achieving $\sim$95\% precision and recall. More granular typing results in worse metrics. In CEERS, we find 9 M8$\pm$2, 2 L6$\pm$2, 1 T4$\pm$2, and 15 T8$\pm$2. We compare the observed long wavelength NIRCam colours —not used in the kNN— to those expected for brown dwarf atmospheric models. The NIRCam F356W-F444W and F410M-F444W colours are redder by a magnitude for the type assigned by the kNN, hinting at a wider variety of atmospheres for these objects.  
We find a 300-350pc scale-height for M6$\pm$2 dwarfs plus a second structural component and a 150-200pc scale-height for T6$\pm$2 type dwarfs, consistent with literature values.

\end{abstract}

% Select between one and six entries from the list of approved keywords.
% Don't make up new ones.

\section{Introduction}

With the successful start of operations of the NASA/ESA/CSA's James Webb Space Telescope (JWST) a new window has opened up on the lowest mass stars of our Milky Way. These near-infrared stellar objects are interlopers in the new, deep extra-galactic observations, but low-mass (sub)stellar members of the Milky Way also encode information on the shape and history of our Galaxy. Here, we report the numbers of such low-mass (sub)stellar objects found in the Cosmic Evolution Early Release Science Survey \citep[CEERS,][]{Finkelstein22} collaborations first NIRCam observations \citep{Bagley22}. 

Counting stars to infer the morphology and scales of the Milky Way is a classic experiment in Astronomy. This was the original motivation behind mapping the night sky, begun initially with photographic plates, moving to wide-field CCDs, and now space-based imaging with missions such as the {\it Hubble} and {\it Webb Space Telescopes}. 

However, counting stars is also an approach that can easily fall into conceptual problems and incomplete information \citep[e.g., Herschel 1785;][]{Kapteyn22}. The disk of the Milky Way was eventually measured and mapped using Giant stars with ever increasing precision \citep{Gilmore83,Gilmore84,Siegel02}. The statistics are now such that individual stellar kinematic populations can be tracked \citep[e.g.][]{Hsu21a} and a link between the scales of the disk and metallicity is evident \citep[e.g.][]{Bovy12}. 

Near-infrared mapping of the sky has revealed dwarf galaxy populations in the local solar neighbourhood \citep{Reid01,Cruz04}. Thanks to the WISE satellite \citep{Wright10}, a near-complete census of the nearest 20 parsecs around our Sun, the numbers of these brown dwarfs are now known, adding a spectral class in the process \citep{Reid08,Kirkpatrick21}.

These lower-mass stars have gained attention as a nuisance population in the search with the Hubble Space Telescope and now James Webb Space Telescope for high-redshift ($z>>6$) galaxies \rev{as both populations overlap in color space} \citep{Caballero08, Wilkins14, Finkelstein15b}. This provides a completely independent scientific motivation to characterize \rev{the number of ultracool dwarfs} in the Milky Way. 

Deep \textit{Hubble} observations have constrained the size of the Milky Way disk thanks to the unambiguous identifications of stars as point sources as even high redshift galaxies are resolved and enough colours or grism spectra exist for type identification. Brown dwarfs have been identified in the Hubble Deep Field \citep{Pirzkal05}, GOODS/CANDELS fields  \citep{Stanway08, Pirzkal09} and individual parallel fields \citep{Ryan05,Ryan11,Holwerda14,van-Vledder16}. But with low statistics and only a few lines of sight out of our Milky Way, these studies remain uncertain on the scale-height, especially as a function of brown dwarf (sub)type.

\cite{Ryan17} introduces a new concept to model the scale-height for different sub-types of brown dwarfs: it implies that as brown dwarf's atmospheres cool \revtwo{(i.e. these objects age and thus become a later type)}, their population scatters and kinematically heats up, resulting in a greater Galactic scale-height. Depending on the star-formation history of the Milky Way, this cooling/aging effect with kinematic heating of the population as a whole, should be clear in the brown dwarf type-scale-height relation. \cite{Ryan22} refined the above model and included the kinematic tracers showing direct evidence for the vertical mixing of this population\cite{Aganze22a,Aganze22b}. 
Additional prior constraints can come from the local IMF \citep{Kirkpatrick21}, local 3D T-dwarf kinematics \citep{Hsu21a}, and the previous determined scale-heights \citep{Aganze22a,Aganze22b} from \revtwo{the WFC3 Infrared Spectroscopic Parallel Survey WISPS} grism observations. \revtwo{These Hubble Space Telescope searches for brown dwarfs are limited to $m_J = 24$AB for pure-parallel photometry and grism observations.}

The challenge is therefore to type M/L/T/Y dwarf objects sufficiently accurately to map their distances photometrically. The broad filters typically employed on JWST/NIRCam deep observations such as CEERS \citep{Finkelstein22} and PEARLS \citep{Windhorst23} programs may hold enough information to do so \citep{Holwerda18}.
\revtwo{The CEERS NIRCam observations were in a large part designed to identify sources for follow-up with the other JWST instruments to characterize their nature \citep[see][]{Bisigello23,Carnall23,Coogan23,Costantin23,Arrabal-Haro23a,Arrabal-Haro23b,Gomez-Guijarro23,Long23,Magnelli23,Shen23,Vega-Ferrero23,Kocevski23b,Larson23}. Our aim here is to evaluate how suitable such JWST/NIRCam observations are for Galactic science, specifically low-mass (sub)stellar objects.}

Here, we explore the use of the colour space of the early CEERS NIRCam observations to identify (sub)stellar objects in these fields. We classify them using the templates of nearby examples using k-nearest neighbours. And finally we constrain the scale-height of the Milky Way using the identified (sub)stellar populations in the CEERS field, and compare these with existing estimates. 

% Meanwhile, all-sky and wide ground based near-infrared surveys have gained in depth and wavelength coverage and two new studies present optical/near-infrared catalogs with photometric sub-typing of a very large part of the sky: \cite{Ahmed19} presents a Northern Sky catalog based on Sloan Digital Sky Survey and the UKIRT Infrared Deep Sky Survey, focusing mostly on M-type dwarfs. \cite{Carnero-Rosell19} present the Dark Energy Survey folding in information from Vista Hemisphere Survey (VHS) DR3 and Wide-field Infrared Survey Explorer (WISE). This Southern sky survey includes all brown dwarf of type M/L/T. 

%The observational situation is that either from the ground or using HST grism observations, the scale-height of the Milky Way disk is relatively well determined for different types of these brown dwarfs. However, both \cite{Holwerda14} and \cite{van-Vledder16} found clear evidence for a second structural component in the BoRG[z8] star counts. 

\section{CEERS}

The Cosmic Evolution Early Release Science Survey \citep[CEERS, PI: S. Finkelstein;][]{Finkelstein17} is a JWST early release science (ERS) project to explore the early Universe and the early evolution of galaxies using a variety of instrument modes, shortly after telescope commissioning \citep{Rigby23,Rigby23a}.

CEERS has covered $\sim$90 sq. arcmin of the EGS field with JWST \citep{Menzel23,McElwain23,Gardner23} imaging (and a portion with spectroscopy) using NIRCam \citep{Rieke23} described in \cite{Bagley23a}, MIRI \citep{Wright23} described in \revtwo{Yang et al.\ \textit{(in preparation})}, and NIRSpec \citep{Jakobsen22,Boker23} described in \revtwo{Arrabal Haro et al.\ \textit{(in preparation)}}. CEERS demonstrates, tests, and validates efficient extragalactic surveys with coordinated, overlapping parallel observations in a field supported by a rich set of HST/CANDELS multi-wavelength imaging \citep{Grogin11,Koekemoer11}. \revtwo{The CEERS NIRCam data were obtained in two epochs, in June and December 2022.}
% The observations, tests, and data products have paved the way for Cycle 2 observations.

The CEERS science goals include the identification of high-redshift galaxies through NIRCam colours \citep{Akins23,Finkelstein22c}, confirming candidate high-redshift sources with NIRSpec or ground-based follow-up
\citep[e.g.][]{Fujimoto22a,Fujimoto23}, and probe the physical conditions in these early galaxies using emission line diagnostics and mid-infrared emission \citep{Arrabal-Haro23a,Arrabal-Haro23b,Jung22,Jung23,Tang23,Zavala23}. The NIRCam filters are sensitive to rest-frame optical emission and thus ideal to probe the evolution of structure starting at a much earlier epoch than ever before \citep{Guo23,Long23,Huertas-Company23a,Magnelli23,Shen23b,Vega-Ferrero23}. These two science topics, the identification for follow-up of high-redshift targets and the structure of earlier epoch galaxies were the main drivers of the observational design. 
The early science methodology and results are presented in a series of Key Papers 
\citep{Finkelstein23,Kocevski23a,Kartaltepe23,Papovich23,Perez-Gonzalez23a,Yang23}.
Key for these science aims is the NIRCam imaging for discovery and characterization of sources in this field. 
However, these CEERS JWST/NIRCam observations have been designed from the start to explore and characterize extragalactic topics and no special consideration was given for Galactic science.

\begin{figure}
    \centering
    \includegraphics[width=0.5\textwidth]{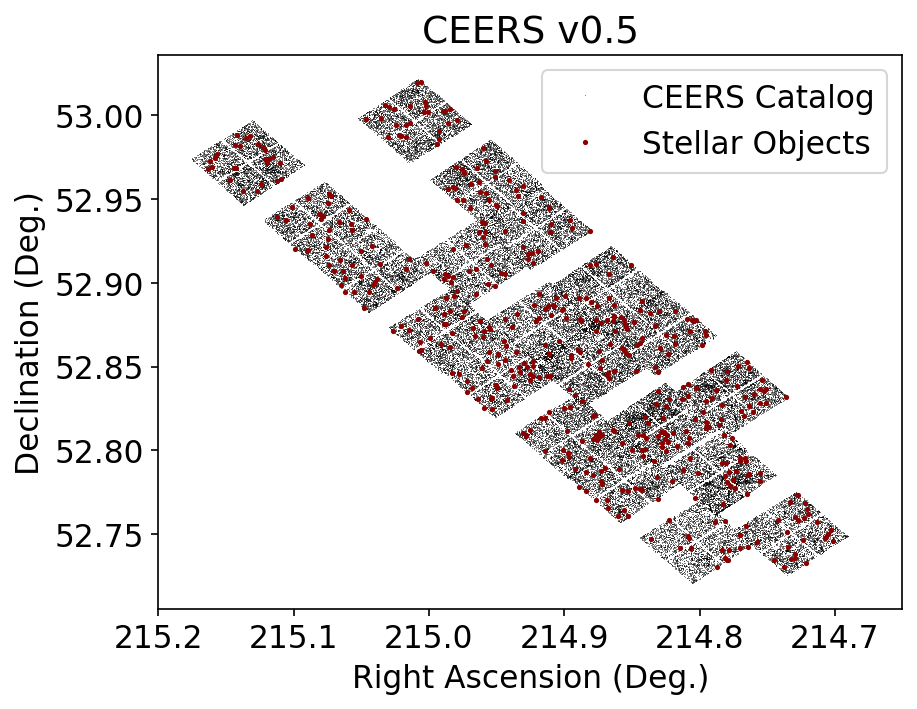}
    \caption{Small black dots denote all objects in the CEERS photometric catalog used for this work.  The red dots denote the positions of stellar objects identified in Figure \ref{f:ceers:mAB-r50}. }
    \label{f:ceers:field}
\end{figure}

\section{Data}
\label{s:data}
% https://jwst-docs.stsci.edu/files/182256933/182256934/1/1669487685625/NRC_ZPs_0995pmap.txt05
The JWST/NIRCam images used are the pixel-aligned mosaics described in \cite{Bagley23a}, v0.6 on the CEERS website. Empirical PSFs were created by stacking stars for the entire mosaic in the case of HST and NIRCam.

We make use of the internal CEERS photometry catalog.  This will be discussed in Finkelstein et al. ({\em in prep.}), but the methodology is summarized here.  The PSF of F606W, F814W, F115W, F150W and F200W is matched to that of F277W (see Table \ref{t:observations}).  For bands with larger PSFs, the redder NIRCam filters, F356W and F444W, and all other HST/WFC3 filters, they calculated correction factors by convolving the shorter wavelength images to the larger PSF of the F444W, and measuring the flux ratio in the native image to that in the convolved image.  By applying this correction factor the fluxes measured in the images with larger PSFs, this corrects for the missing flux under the assumption that the morphology does not change significantly. This is especially useful for small high-redshift sources or faint stellar photometry as it is used here. The photometry using Source Extractor \citep[v2.25.0][]{SE,seman} is described in Finkelstein et al ({\em in prep}).

%Photometry was performed with Source Extractor \citep[v2.25.0][]{SE,seman} in two image mode, using a weighted sum of the native resolution F277W and F356W as the detection image. For the detection image, the CEERS team used 1/sqrt(WHT) map with map\_rms weighting, such that Poisson noise (which is in the ERR maps) did not affect the detections.  These effective ERR maps were scaled to have a median value matching that of the ERR map, similar to the approach of \cite{Trenti11}.  For the measurement/photometry image, we used the relevant photometry filter's ERR map.  This current catalog has DETECT\_THRESH=1.4 and DETECT\_MINAREA=5, derived iteratively via visual inspection to minimize spurious sources while maximizing completeness.  We currently use this same threshold for all fields.

The fits catalog contains photometry in units of nJy. We made use of the fiducial flux and flux error columns, which were corrected to total, and with flux errors calculated empirically from the images. 
% The "FLUX\_FILTER" and "FLUXERR\_FILTER" columns contain our fiducial total fluxes.  
Fluxes were converted to AB magnitudes using a conversion of $\rm m_{AB} (FILTER) = -2.5 \times log_{10}(FLUX\_FILTER)+ 31.4 $. 
% Here we use the v051 version of the CEERS catalogs on ten NIRCam fields in the CEERS program. \rev{We opt for the use of this catalog for ease of use and the ability to directly compare to other CEERS results on colours of dusty dwarf galaxies at higher redshift \citep{Bisigello23} and \revtwo{intermediate} redshift sources \citep{Perez-Gonzalez23a,Perez-Gonzalez23}. }
The F277W+F356W  image was the detection image with the other filters run in dual-image mode to obtain photometry \citep[see][Finkelstein {\it in preparation} for details]{Finkelstein22}. The half-light radius from Source Extractor for the F200W filter image is included in this catalogue. The position of all the sources in the catalogue is shown in Figure \ref{f:ceers:field}.

\begin{figure}
    \centering
    \includegraphics[width=0.5\textwidth]{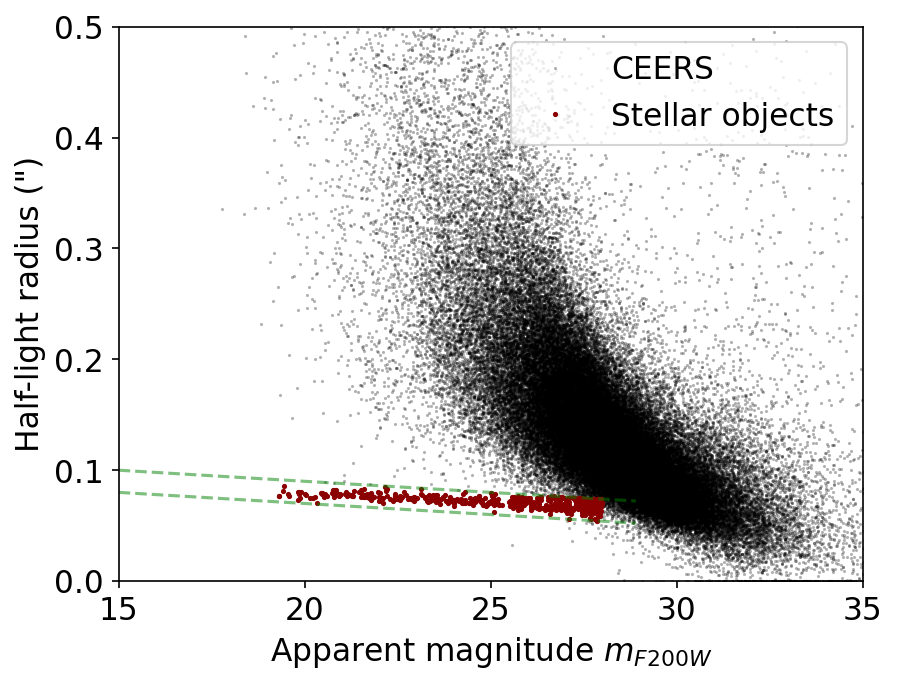}
    \caption{The apparent magnitude of objects in the CEERS photometric catalogue in the F200W detection filter (AB magnitude) and the effective radius computed by Source Extractor in the same filter. The green dashed lines delineate the locus of unresolved sources. We identify the red objects as unresolved stellar candidates. }
    \label{f:ceers:mAB-r50}
\end{figure}

\subsection{Stellar Catalog}

We use the effective radius measured in the F200W image of the two-image Source Extractor run described above to identify unresolved objects in the stellar locus. Figure \ref{f:ceers:mAB-r50} shows the effective radius as a function of the F200W AB magnitude. The green dashed lines show the limits of the stellar locus, similar to the approach in \cite{Ryan11} and \cite{Holwerda14}. The unresolved stars' half-light radius is less than 0\farcs1 and declines slightly with fainter objects. \revtwo{This is a known effect in HST imaging because the growth curve from which Source Extractor derives the half-light radius loses the lower surface brightness pixels which make a larger fraction for the flux for fainter sources.}
The red marks in Figure \ref{f:ceers:mAB-r50} are the stellar candidates marked in Figure \ref{f:ceers:field} as well. The utility of the half-light radius measured by Source Extractor translates well from HST to JWST with a slightly different dependence between the half-light radius and magnitude. We empirically determine that this stellar identification works well to m$_{AB}\sim28$ for the CEERS field. 

We use AB magnitudes derived to determine colours for the three feature spaces, JWST/NIRCam, HST+JWST combined, or either ranked by filter wavelength or colours based on only HST and JWST filter pairs.

\begin{figure*}
    \centering
    \includegraphics[width=0.49\textwidth]{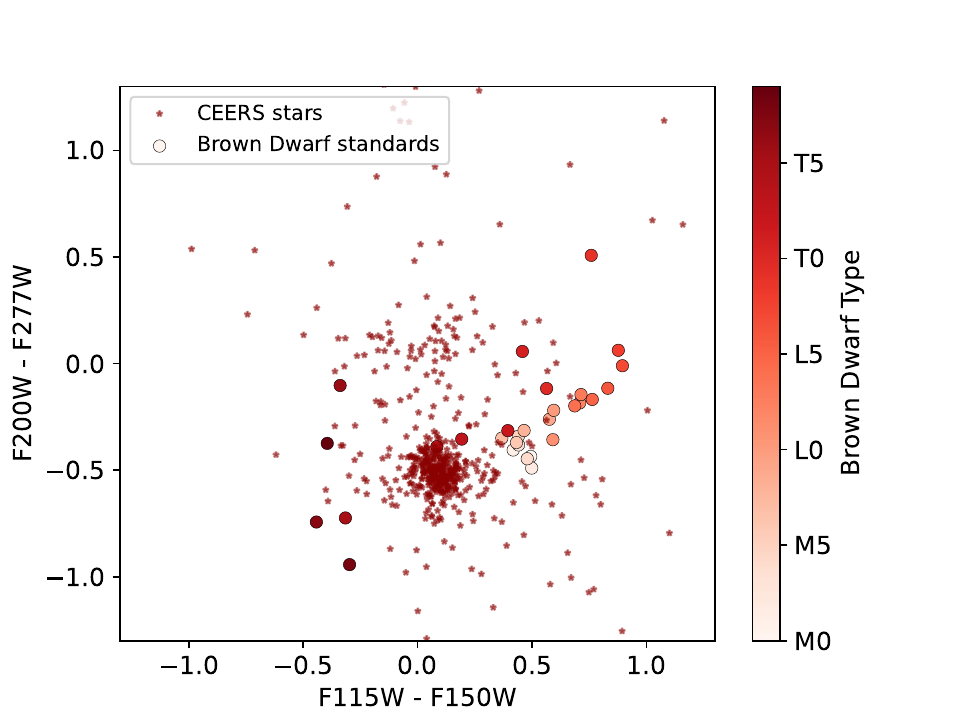}    \includegraphics[width=0.49\textwidth]{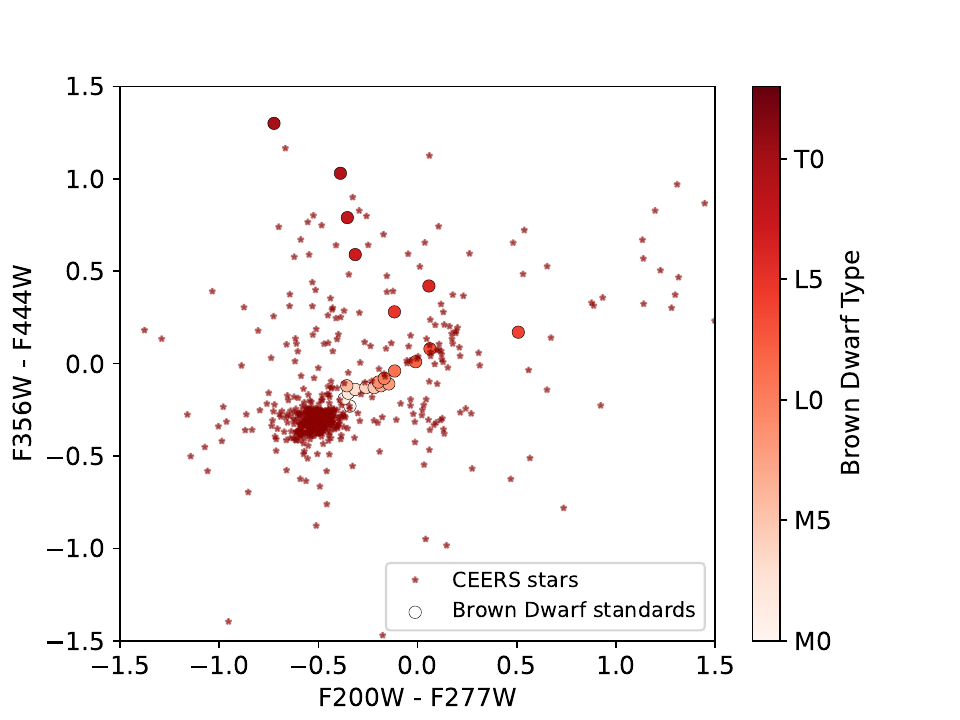}
    \caption{The 
    F115W-F150 and F200W-F277W,  
    F200W-F277W and F356W-F444W  colour-colour space for the stellar objects identified in Figure \ref{f:ceers:mAB-r50} with the brown dwarf standard stars overplot as tracks in increasing type. With the intrinsic variation in colour with brown dwarf type, this shows it is plausible a majority of the stellar sources are possibly brown dwarf type objects. No F410M filter information is available for the standard brown dwarfs.}
    \label{f:ceers:NIRCam:color-color-space}
\end{figure*}

Figure \ref{f:ceers:NIRCam:color-color-space} shows the NIRCam colours space with the SpeX Prism Library Analysis Toolkit (SPLAT) template stellar colours overplot of increasing type \citep{Burgasser17,Holwerda18}. A majority of the stellar objects have colours similar to brown dwarf templates.

\begin{figure}
    \centering
    \includegraphics[width=0.5\textwidth]{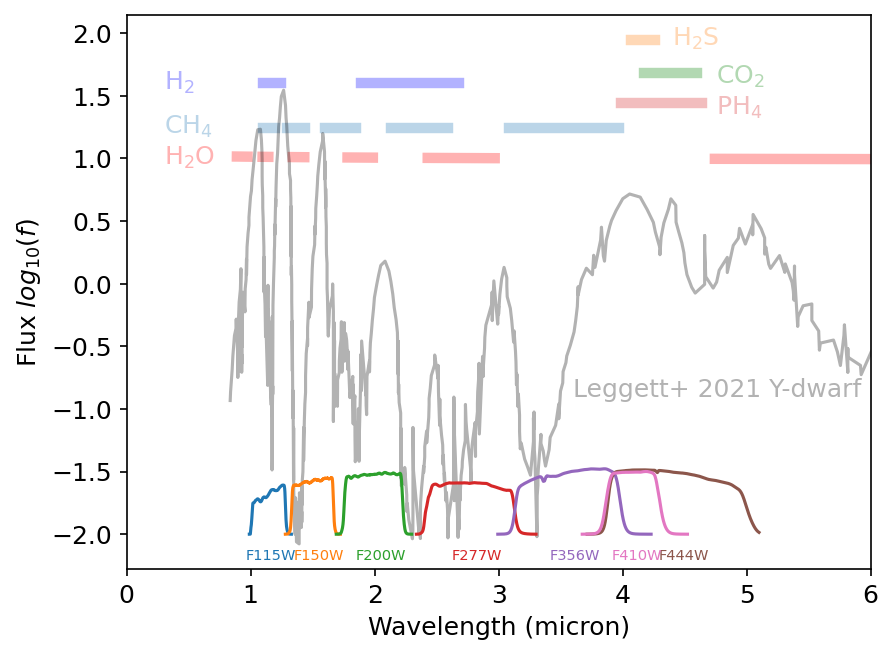}
    \caption{An example of a Y-dwarf from \protect\cite{Leggett21} with some of the absorbing molecule species marked and the CEERS NIRCam filters used annotated at the bottom. The JWST/NIRCam blue filters (F115W, F150W, F200W and F277W) are sensitive to the molecular absorption features that define the classes. The longer wavelength filters (F356W, F410M, and F444W) are more sensitive to overall temperature and different molecular absorption features.   }
    \label{f:ydwarf}
\end{figure}
\section{Classifying M-l-T-Y type dwarfs with NIRCam colours }
\label{s:classifying}

A problem in identifying M-L-T-Y dwarf stars from just photometry is that, while there are some clear absorption features in their spectra in the 0.5-2 micron range, these easily become a somewhat degenerate space when observed with wide filters of current and near-future near-infrared imaging space observatories \citep{Holwerda18}.
\revtwo{Figure \ref{f:ydwarf} shows an example of a Y-dwarf spectrum from \cite{Leggett21} with the CEERS NIRCam filter curves to illustrate. Depending on the atmospheric composition of the brown dwarf, one can expect very different NIRCam colours.}

The challenge is now to classify as well as possible with just NIRCam broad colours. Previous work \citep{Ryan11,Holwerda14} solved this with colour-colour spaces defined in HST filters. The issue is that these are hard cuts in colour space and the problem of broad classification is not as clear-cut as that.

Here, we use the JWST/NIRCam colour predictions based on SPLAT spectral library from \cite{Holwerda18} to classify unresolved sources in CEERS.
To do so, we tried k-means Nearest neighbour (kNN), a support vector machine (SVM) and a random forest (RF, an average of decision trees) on the bluest colours. Because the SPLAT spectra only extend to 2.5$\mu m$, the longest NIRCam wavelengths cannot be included for the training of the ML algorithms but we can compare the resulting type predictions to the standard stars for each type, for which there is analogue WISE satellite W1-W2 colour available. 

\rev{Because the colour space of brown dwarfs represents a space with steep gradients and a mixture of physical parameters that drive the observed NIR spectra, we prefer this approach over fitting templates to each SED with assuming the lowest $\chi^2$ is the optimal fit. By using SPLAT observed spectra rather than model templates, the kNN fit allows for observed but still unexplained variance in each type. }

\subsection{Training Sample}
\label{s:training}

The full JWST/NIRCam colour sample available based on SPLAT spectroscopy \citep{Burgasser17} is described in \cite{Holwerda18}. Briefly, there is JWST/NIRCam photometry estimates for the \rev{30} standard stars which define the spectral type and for all the SPLAT spectra which capture the intermediate types and variance within spectral types. We use the latter as a training set. The photometry from \cite{Holwerda18} is in Vega magnitudes and we convert to AB magnitudes first. There is NIR colours in both HST/WFC3 and JWST/NIRCam filters. We use three \rev{colour} feature spaces, a NIRCam-only, a NIRCam and HST one with colours separated by instrument, and a third where colours use filters in wavelength order, mixing HST and JWST filters. 

\begin{figure}
    \centering
    \includegraphics[width=0.5\textwidth]{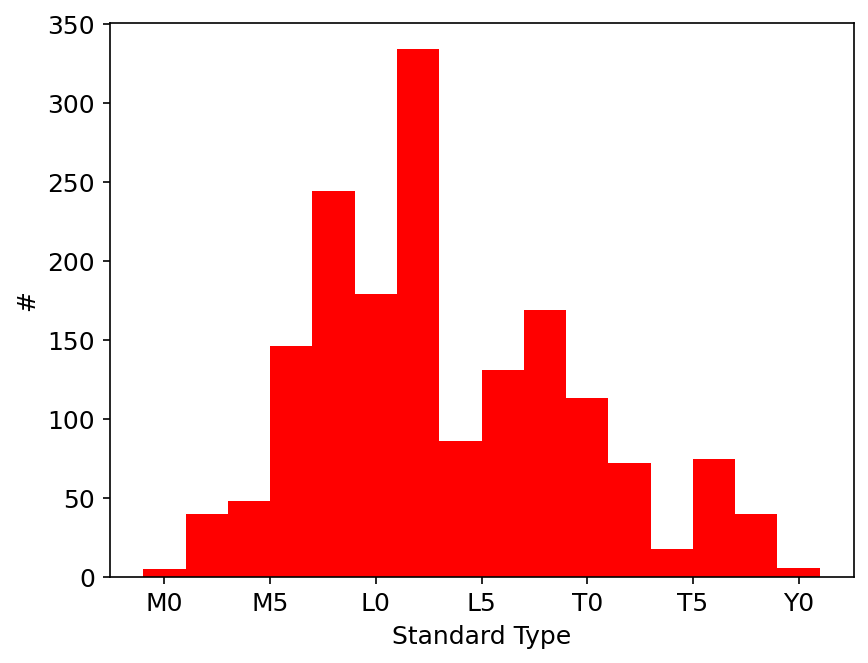}
    \caption{The distribution of standard types for the SPLAT training set. The training set is dominated by late M- and early L-dwarfs. }
    \label{f:splat:training:types}
\end{figure}

For the labels one can employ index-type or standard\_type from the SPLAT catalog \citep{Burgasser17}, the \revtwo{latter} a classification by the spectral index and the former the class to which standard star they resemble most. The indextype has fewer stars classified as intermediate than the standard\_type does. We adopt the standard\_type throughout. 
We assign a numerical type as follows: M-types are 0-0.9, L-types 1.0-1.9, T-types 2.0-2.9 and Y-types 3 and above. Later, when applying these labels to the kNN multi-classing, we multiply these by 10 because the kNN can only handle integer classes. Late M- and early L-types \rev{are much more represented in the SPLAT training set compared to the later types because their relative luminosity. But all types and subtypes are present in the training set with reasonable sampling (Figure \ref{f:splat:training:types}). }

\begin{figure}
    \centering
    \includegraphics[width=0.5\textwidth]{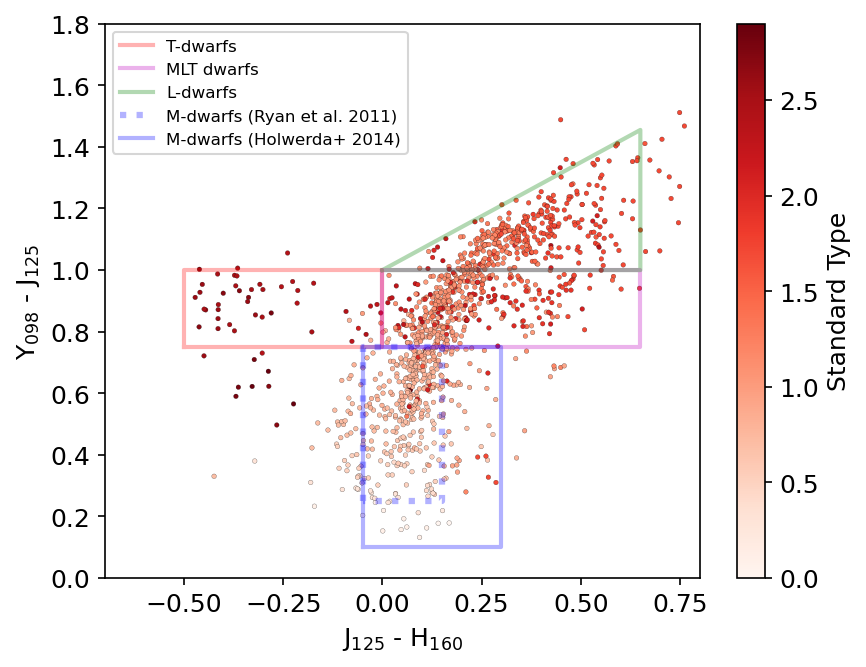}
    \caption{The distribution of HST/WFC3 colours for the standard types for the training set. The colour-colour selections from \protect\cite{Ryan11} and \protect\cite{Holwerda14} for M-, L-, and T-dwarfs are indicated with coloured boxes. This illustrates the issue with hard cuts: in a two-dimensional space like this, there is still ample mixing of types. A photometric classifier based on proximity in this space is expected to perform better than these initial colour-colour cuts.}
    \label{f:splat:hst:cc}
\end{figure}

\cite{Ryan11} and \cite{Holwerda14} used the near-infrared colour-colour space to broadly type the M-L-T dwarfs in parallel HST/WFC3 fields. Figure \ref{f:splat:hst:cc} shows the original HST/WFC3 colour-colour space used for broad photometric typing of these objects. Broad classes are still mixed in this colour space and subtyping \revtwo{using hard cuts in colour-colour space} is not possible. 

\begin{figure*}
    \centering
    \includegraphics[width=\textwidth]{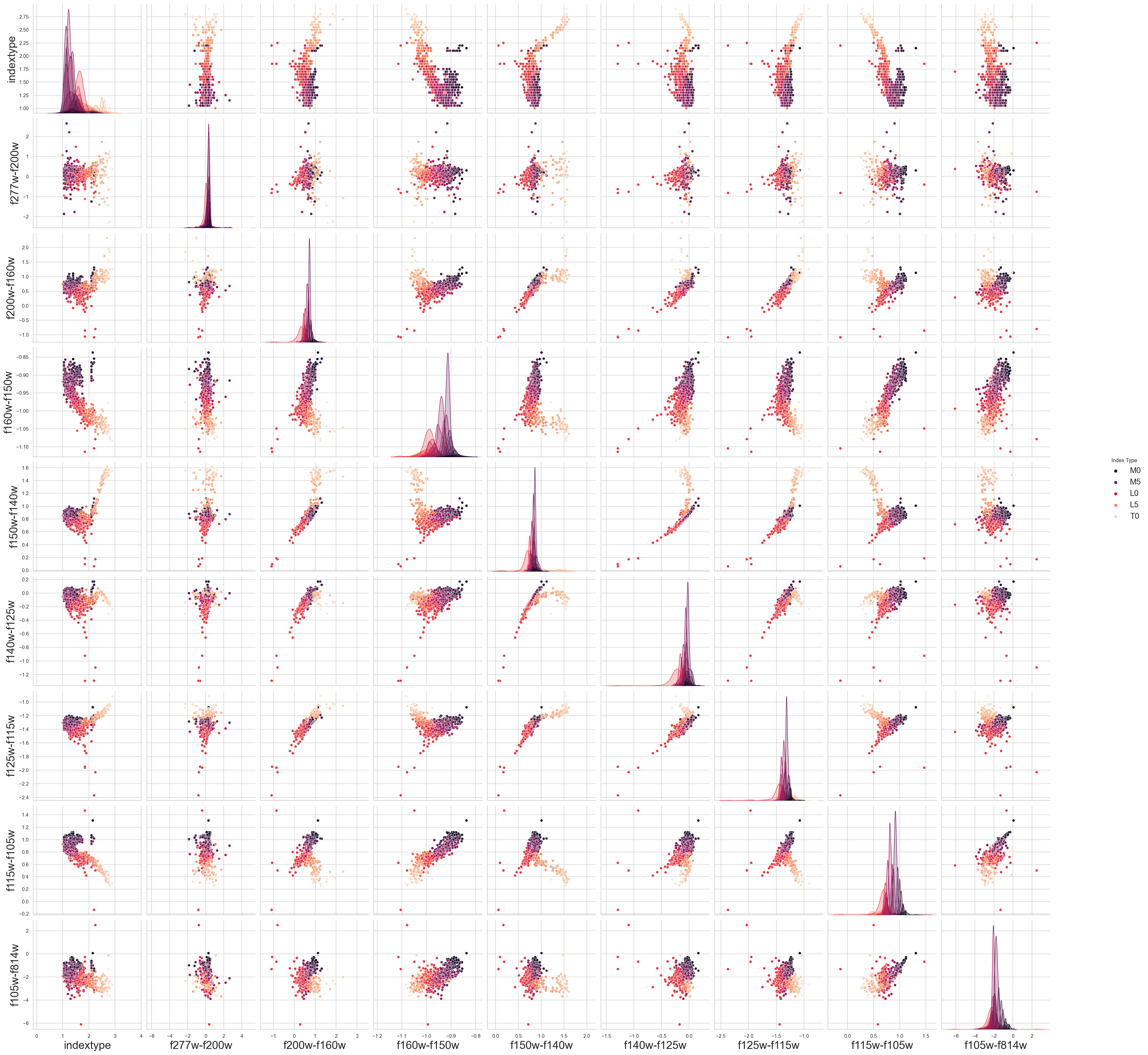}
    \caption{The full feature space of the photometric catalog generated from SPLAT spectra of nearby brown dwarfs for the JWST/NIRCam filters in CEERS. Some parts of the colour-colour space are well-suited to broadly classify these objects but the space is degenerate, an ideal scenario to maximize return with a machine learning approach.}
    \label{f:splat:IDE:jwst-only}
\end{figure*}

Figure \ref{f:splat:IDE:jwst-only} shows a NIRCam equivalent of a colour-colour plot of the SPLAT catalog. A similar separation of objects could conceivably be used. Rather than picking a particular colour-colour combination for broad classification and hoping to sub-class these objects with the remaining information, we opt to subclass directly, varying the \rev{spectral subtype} resolving power ($\Delta T$).
As said above, we use three feature spaces: the bluer NIRCam colours, and two iterations of the CEERS HST+JWST complete photometry, one using only instrument specific colours (inst), i.e. only combining HST filters into a colour, and one with the colours arranged by wavelength (wav). For example, in the latter, we combine HST's F160W and NIRCam's F150W into a single colour. 

\begin{figure*}
    \centering
    \includegraphics[width=\textwidth]{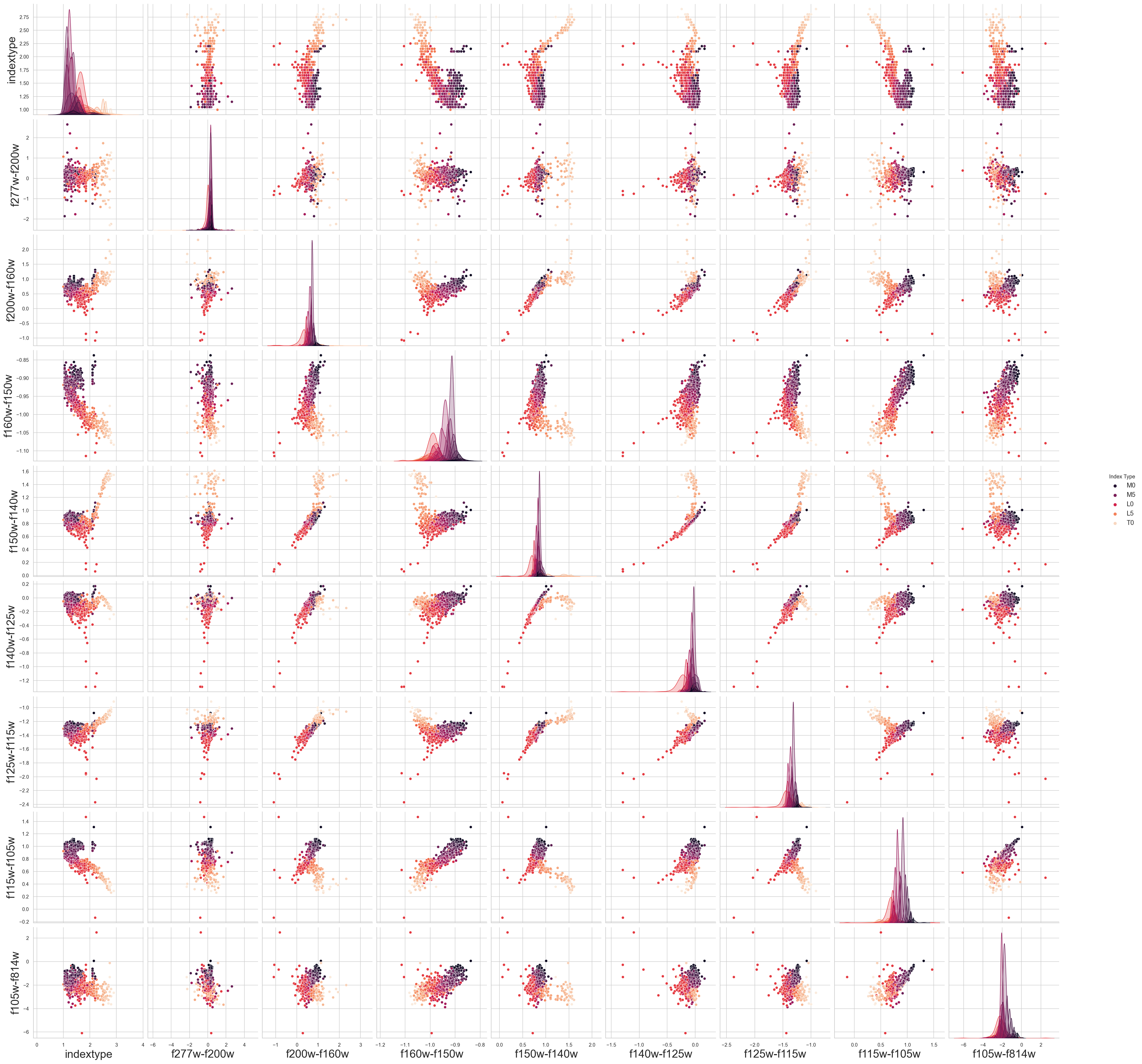}
    \caption{The full feature space of the photometric catalog generated from SPLAT spectra of nearby brown dwarfs for the HST/WFC3 filters in CEERS. Some parts of the colour-colour space are well-suited to broadly classify these objects but the space is degenerate, an ideal scenario to maximize return with a machine learning approach.}
    \label{f:splat:IDE:hst-only}
\end{figure*}

\begin{figure*}
    \centering
    \includegraphics[width=\textwidth]{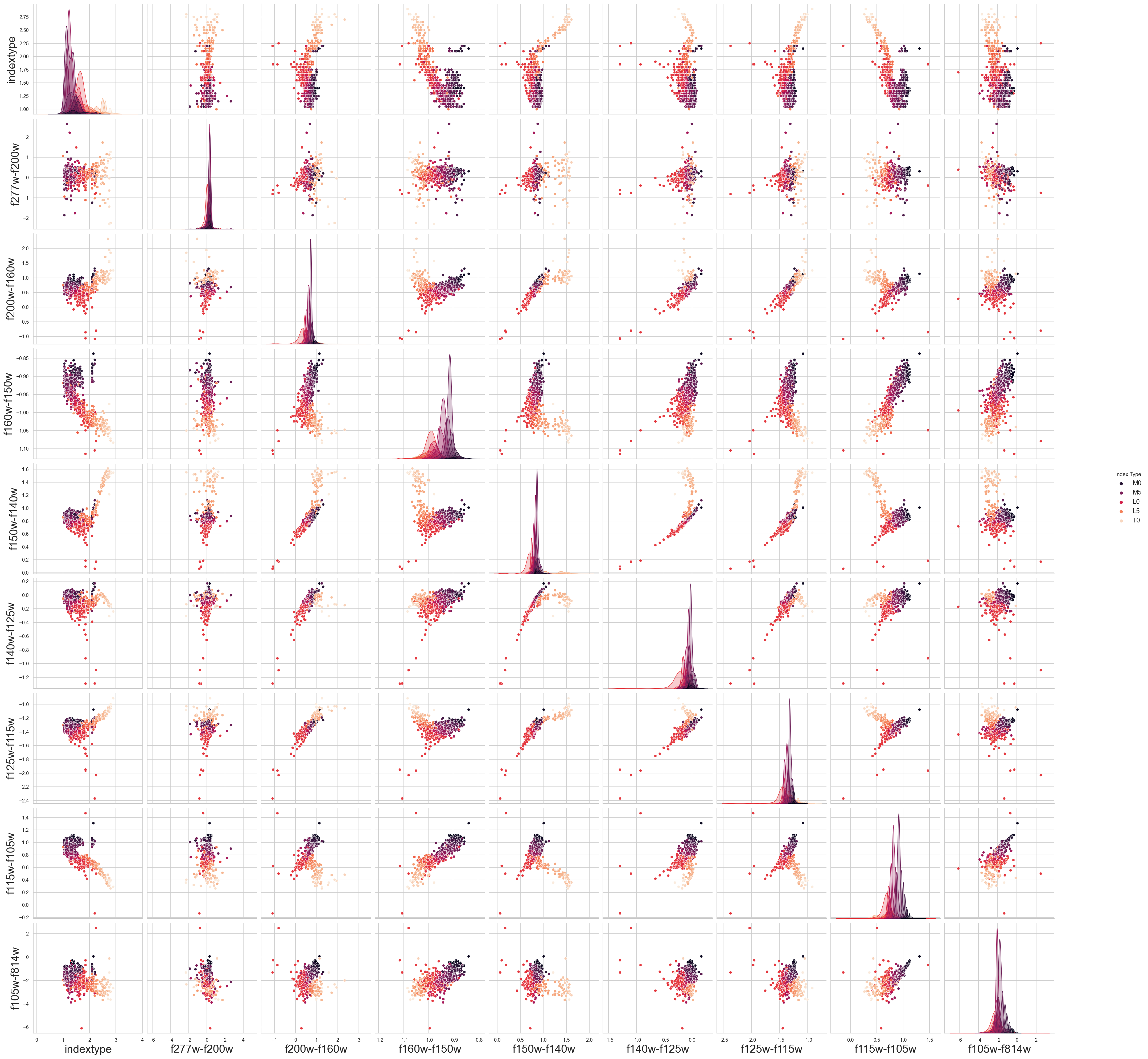}
    \caption{The full JWST+HST feature space of the photometric catalog generated from SPLAT spectra of nearby brown dwarfs for the HST/WFC3 and JWST/NIRCam filters in the CEERS catalog, sorted by wavelength. This extensive colour feature space may well be suited to classify sources in deep HST+JWST data but mixing instruments for the colours may be susceptible to systematics.}
    \label{f:splat:IDE:hst+jwst}
\end{figure*}

Figures  \ref{f:splat:IDE:jwst-only}, \ref{f:splat:IDE:hst-only} and \ref{f:splat:IDE:hst+jwst} show the parameter space and the classification into broad types as the colour-coding. Colour-colour space has already proven itself to work well on broad type (e.g. M, L, T or Y?) but now we will attempt to refine the classification down to subclasses.
 
We start to classify based on 1271 SPLAT \citep{Burgasser17,Holwerda18} spectra from which photometry can be derived in \cite{Holwerda18} with spectral classifications (standard type). There are 40 classes in this training set. 

\subsection{kNN Classifications}
\label{s:knn}

K-Nearest Neighbour classifies objects using the N nearest neighbours in the feature space. Historically, kNN classifies in to two classes (0 or 1). MultiClass classification can be defined as the classifying instances into one of three or more classes. Multiclassing is possible with the {\sc sklearn} implementation of kNN and this is what we employ here. \rev{kNN classification has successfully been used on brown dwarf photometry in \cite{Marengo09}.}
Splitting the SPLAT sample into the typical 80/20 per cent split between training and test set, we classify into the 40 classes of standard type. We initially use the HST+JWST(wav) feature space where colours are computed going from longest to shortest wavelength of the filter, regardless of instrument used. 

We want to know a few different classification specific questions:
(1) How many neighbours are optimal for classification in this feature space?
(2) To what degree of accuracy can one classify within 0.1 of a subtype, 0.2, 0.5 or just the full type? 
(3) How well does it still perform if only JWST or HST information is used?

\subsection{Number of Neighbours}

\begin{figure}
    \centering
    \includegraphics[width=0.5\textwidth]{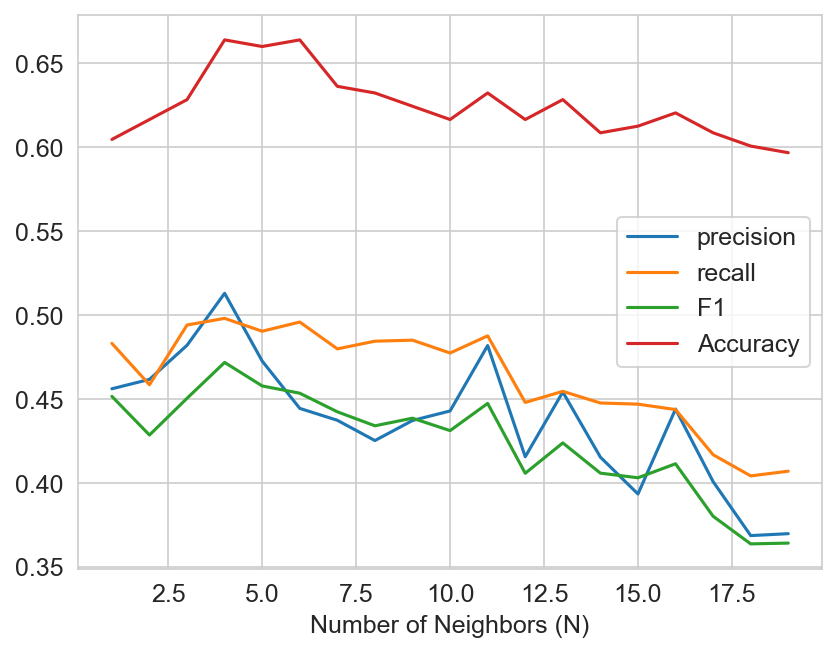}
    \caption{The precision, recall, \rev{F1,} and accuracy\rev{, each expressed as a dimensionless fraction,} as a function of number of neighbours (n) of the test set of the kNN classifier on the SPLAT photometry using the HST+JWST(wav) feature space.}
    \label{f:splat:knn:n-metrics}
\end{figure}

Figure \ref{f:splat:knn:n-metrics} shows the dependence of the kNN performance trained on the full HST+JWST(wav) feature space as a function of the number of neighbours used, the key hyperparameter for kNN classifiers. \rev{Metrics are calculated based on the number of true and false positives and negatives (TP, FP, TN and FN) in case of two classes. Our metrics are accuracy (TP+TN / TP+TN+FP+FN), precision (TP/TP+FP), recall (TP/TP+FN), and F1 (2$\times$ precision $\times$ recall / precision+recall). A good classifier strikes a balance between accuracy (fraction of accurate classifications), the accuracy of positive identification (precision), and the completeness of the identified sample (recall) or a metric of the balance of the previous two (F1). }
The optimal performance in terms of accuracy occurs between n=3 and n=7. Precision and F1 peak at n=4. 
We use n=4 for the remainder of our applications of kNN on these data-sets as the optimal number of neighbours to determine subtype with kNN. \rev{This is comprable to k=3 adopted for Spitzer colour classification adopted by \cite{Marengo09}.}

\subsection{Performance and type accuracy}
\begin{figure}
    \centering
    \includegraphics[width=0.5\textwidth]{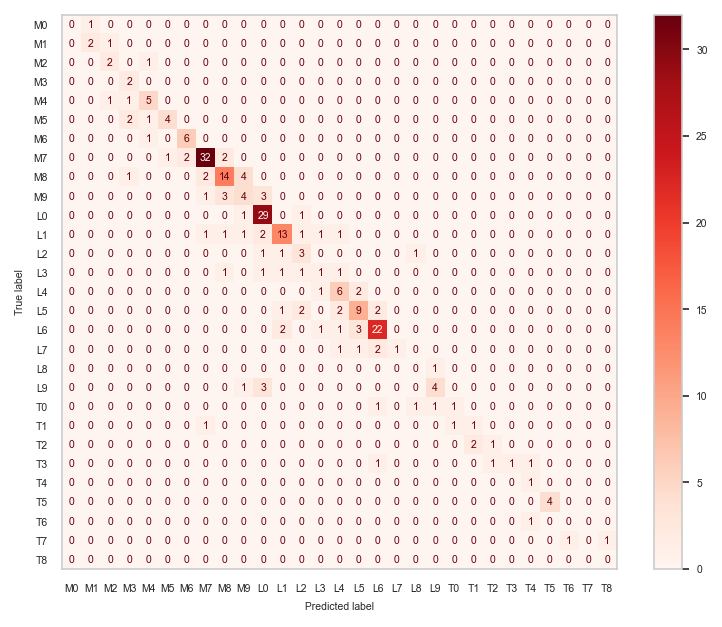}
    \caption{The confusion matrix of the kNN classification into individual subtypes of M, L, T-dwarfs using the HST+JWST(wav) feature space shown in Figure \ref{f:splat:IDE:hst+jwst}.
    The overall metrics can be found in Table \ref{t:knn:metrics} for a type resolution of $\Delta T = 0.1$}
    \label{f:CM:sttype:full}
\end{figure}
% Precision: 0.5131043958630166
% Recall: 0.4981727776216863
% F1: 0.4981727776216863
% Accuracy: 0.6640316205533597
Figure \ref{f:CM:sttype:full} shows the confusion matrix over all 40 types of standard type classes in the training set $\Delta T = 0.1$, the highest type accuracy possible. Accuracy was 66\% and the other metrics are in Table \ref{t:knn:metrics}. The same feature space using the indextype as the label performed notably worse. This level of performance is not useful in classification but the confusion matrix in Figure \ref{f:CM:sttype:full} is encouraging in that it may improve with a somewhat coarser type resolution.

\begin{table}
    \centering
    \begin{tabular}{l l l l l}
Instance                & Accuracy & Precision & Recall    & F1     \\ 
kNN $\Delta T$          & (\%)      & (\%)      & (\%)  & (\%) \\
\hline
\hline
    % & 51        & 49        & 49    & 66 \\

% kNN $\Delta T = 0.1$ & 0.56 & 0.59 & 0.57 & 337 \\
% kNN $\Delta T = 0.2$ & 0.56 & 0.59 & 0.57 & 337 \\
% kNN $\Delta T = 0.3$ & 0.49 & 0.51 & 0.49 & 337 \\
% kNN $\Delta T = 0.3$ & 0.74 & 0.75 & 0.73 & 337 \\
% kNN $\Delta T = 0.5$ & 0.95 & 0.96 & 0.95 & 337 \\ 
% kNN full &    0.85 & 0.85 & 0.85 & 337 \\

0.1 & 0.60 & 0.46 & 0.49 & 0.46 \\
 0.2 & 0.77 & 0.62 & 0.62 & 0.61 \\
 0.3 & 0.77 & 0.78 & 0.79 & 0.78 \\
 0.4 & 0.83 & 0.83 & 0.82 & 0.82 \\
 0.5 & 0.83 & 0.70 & 0.70 & 0.70 \\
 0.6 & 0.83 & 0.68 & 0.68 & 0.68 \\
 0.7 & 0.86 & 0.82 & 0.82 & 0.82 \\
 0.8 & 0.91 & 0.69 & 0.71 & 0.69 \\
 0.9 & 0.89 & 0.88 & 0.91 & 0.89 \\
 1.0 & 0.89 & 0.92 & 0.93 & 0.92 \\

\hline
    \end{tabular}
    \caption{The metrics for the different binning in type for the kNN algorithm. An optimal resolution is $\Delta T = 0.5$ for this colour-space.}
    \label{t:knn:metrics}
\end{table}

\begin{figure}
    \centering
    \includegraphics[width=0.5\textwidth]{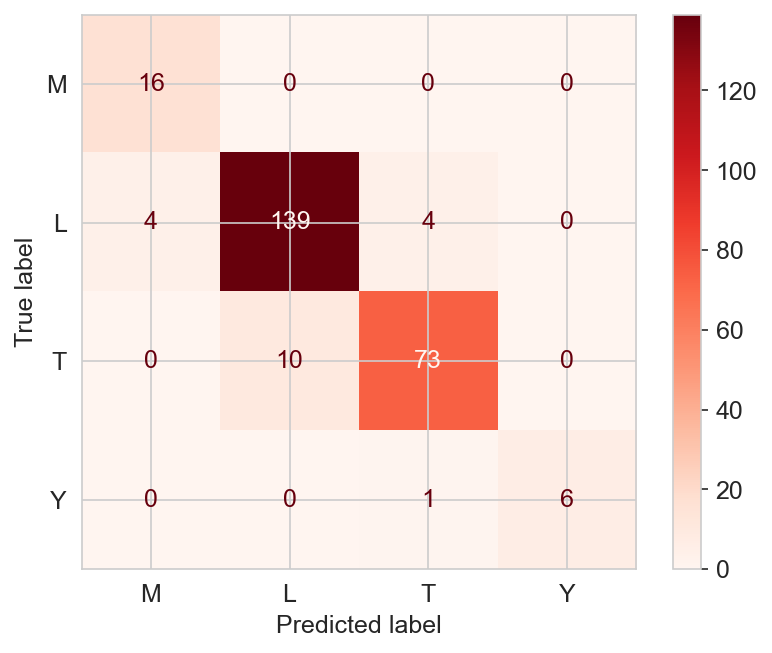}
    \caption{The broad \rev{($\Delta T = 1$)} kNN classification confusion matrix}
    \label{f:CM:sttype:broad}
\end{figure}
Secondly, we explore if just broad class (M/L/T or 0/1/2) classifications can be done with this feature space. Accuracy dramatically improves to 85\% with equally good precision and recall scores. These values were even higher for indextype, probably because it does not contain M-type which is somewhat featureless in the NIR \revtwo{at the spectral resolution of these filters.} Figure \ref{f:CM:sttype:broad} shows the confusion matrix for standard type class for just the broad class for classification.
Broad types one can do very well with these four broad JWST/NIRCam filters. There is a slight tendency of kNN algorithm to class brown dwarfs as later type (below the diagonal). 
\begin{figure}
    \centering
    \includegraphics[width=0.5\textwidth]{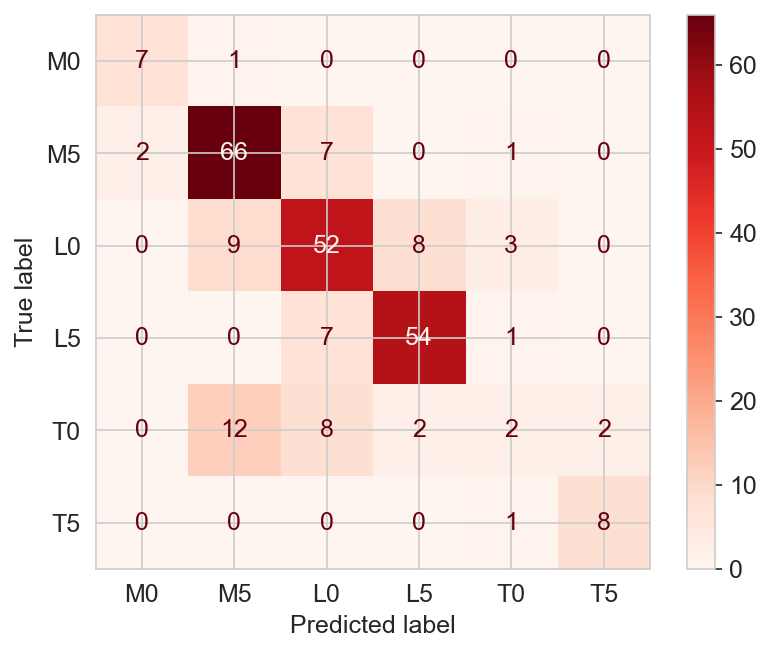}
    \caption{The kNN classification confusion matrix on the HST+JWST(wav) feature space with a much lower type resolution of $\Delta T = 0.5$}
    \label{f:CM:sttype:t05}
\end{figure}

Our question is now how much better the classification can do than broad class? 
First, we classify within half a class, within $\Delta T = 0.5$, where the integer is the broad M, L or T class and then we shall explore 0.4 and 0.2 of a class. Because kNN algorithm does not allow for fractional classes, we multiply the class by 10 (e.g. L1 thus becomes 11 and we divide by 10 afterwards).  
\begin{figure}
    \centering
    \includegraphics[width=0.5\textwidth]{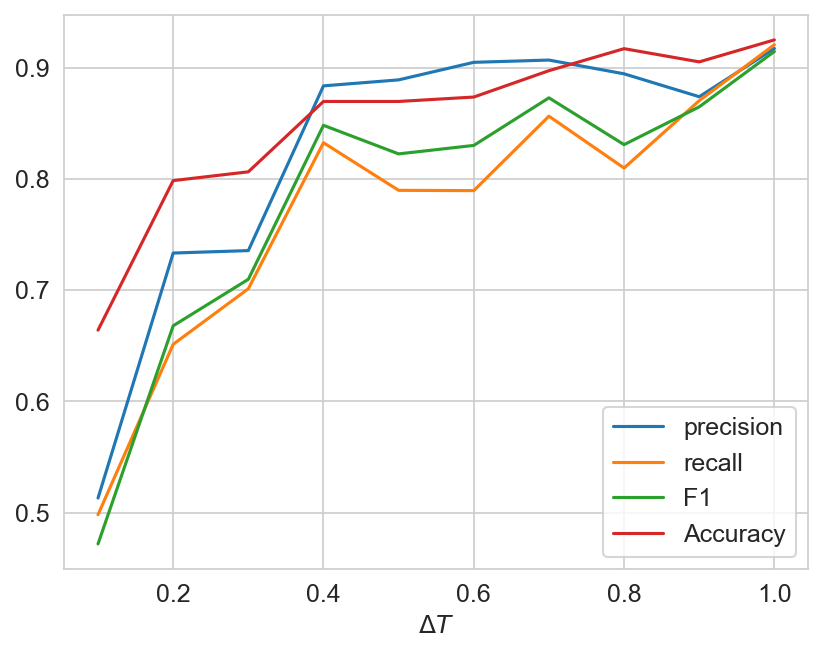}
    \caption{The metrics of the kNN classification\rev{, each expressed as a dimensionless fraction,} as a function of the type resolution ($\Delta T$) for the full HST+JWST/NIRCam. All metrics climb steadily with coarser type resolution. If one wants to classify within a type ($\Delta T = 0.5$), the metrics are $\sim85$\%. Notably, $\Delta T = 0.3$ would perform almost equally well. }
    \label{f:splat:all:DeltaT-metric}
\end{figure}

\rev{Figures \ref{f:CM:sttype:broad} and \ref{f:CM:sttype:t05} show the confusion matrices for the different resolutions in spectral type for the test set. } Figure \ref{f:splat:all:DeltaT-metric} 
shows the metrics of the test set for kNNs trained with different spectral type resolution and these metrics for each kNN run are listed in Table \ref{t:knn:metrics}. 

\begin{figure}
    \centering
    \includegraphics[width=0.5\textwidth]{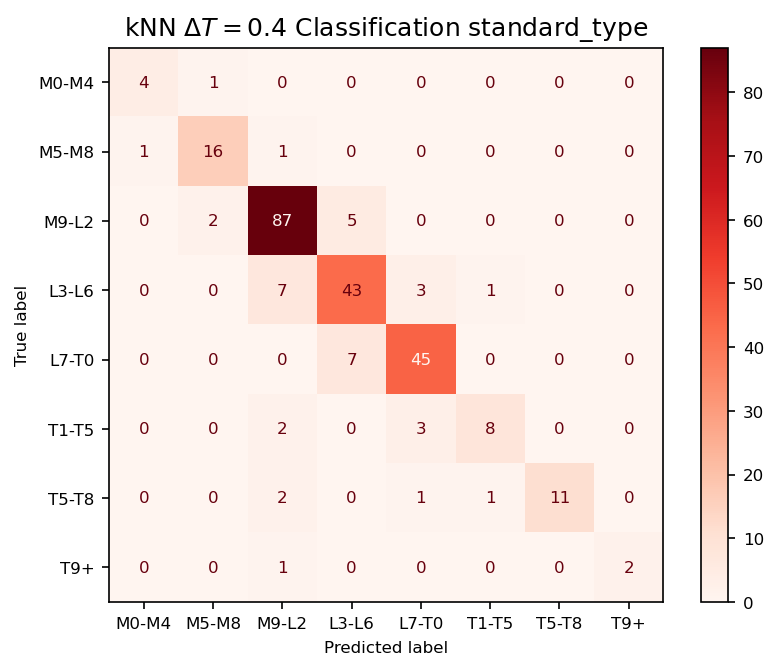}
    \caption{The kNN classification confusion matrix on the HST+JWST(wav) feature space with the optimal type resolution of $\Delta T = 0.4$. This is the kNN classifier used on the CEERS photometry}
    \label{f:CM:sttype:t04}
\end{figure}

\begin{figure*}
    \centering
    \includegraphics[width=0.49\textwidth]{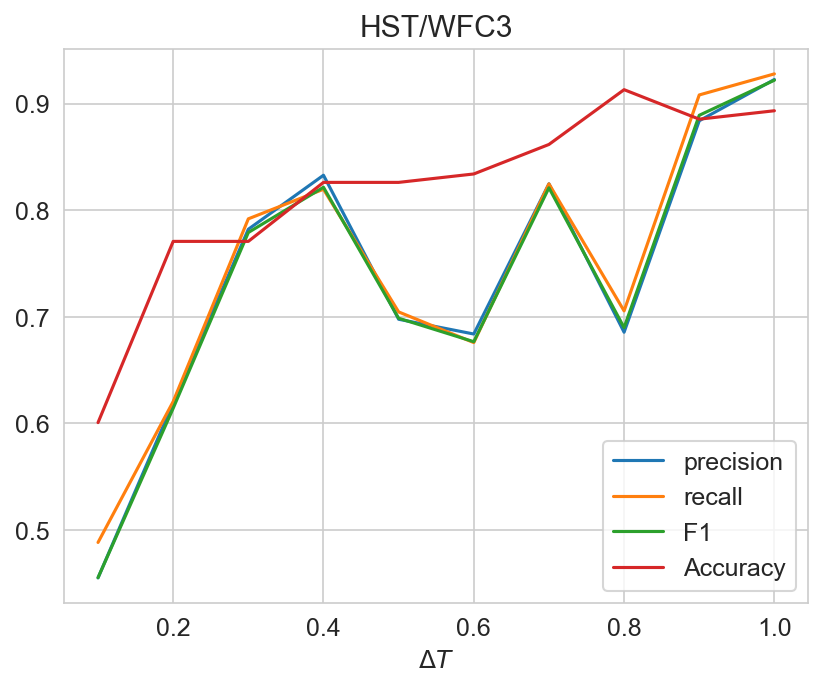}
    \includegraphics[width=0.49\textwidth]{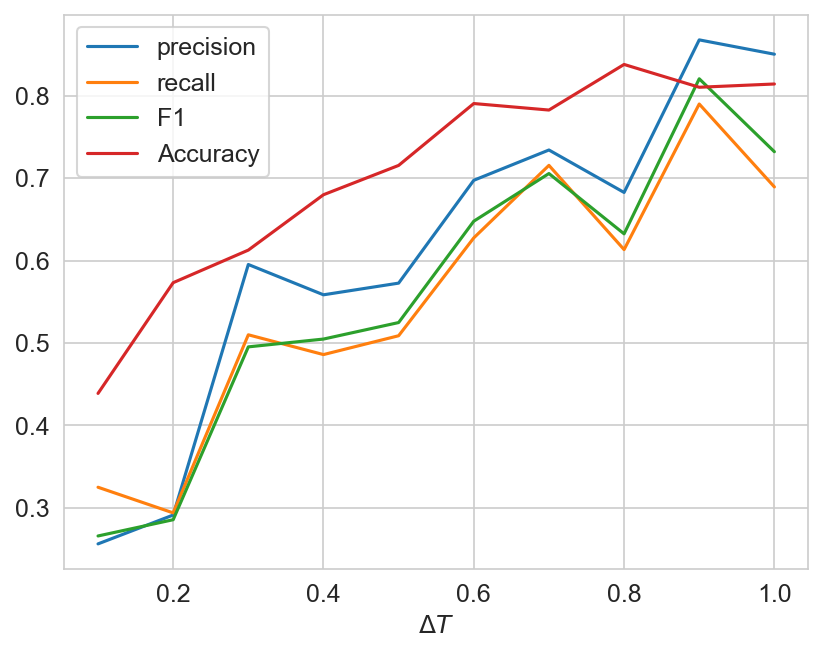}
    \caption{The metrics of the kNN classification\rev{, each expressed as a dimensionless fraction,} as a function of the type resolution ($\Delta T$) for the HST-only filter colour-space (left) and the JWST-only filter colour-space (right). For HST, all metrics climb steadily with coarser type resolution. If one wants to classify within a type ($\Delta T = 0.5$), the metrics are $\sim85$\%. Notably, $\Delta T = 0.4$ would perform almost equally well. For JWST-only, the blue NIRCam filers perform poorer than in combination with HST (Fig. \ref{f:splat:all:DeltaT-metric}) or HST-only (left) until almost full type resolution $\Delta T \sim 1$}
    \label{f:splat:hst:jwst:DeltaT-metric}
\end{figure*}

\rev{
Figure \ref{f:CM:sttype:t04} shows the optimal type resolution of $\Delta T=4$ with k=4. This is the optimum we can achieve with the mix of filters available in CEERS. We note that HST or JWST filter combinations alone also perform optimally with this spectral type resolution but with lower overall metrics (Figure \ref{f:splat:hst:jwst:DeltaT-metric}). }

We varied the number of neighbours for each classification scheme (e.g. \rev{n=2 to n=20}) but this made no real difference in the metrics. \rev{Just under h}alf a type \rev{($\Delta T =0.4$)} clearly is the optimal resolution for the HST three-colour feature space to classify in. This is a substantial improvement the hard colour cuts into broad type by \cite{Ryan11} and \cite{Holwerda14} but falls just short of grism spectroscopy classification which can discriminate between subtypes \citep{Aganze22a,Aganze22b}.

Figure \ref{f:splat:hst:jwst:DeltaT-metric} shows the type resolution of HST-only and JWST/NIRCam-only colour feature spaces. Surprisingly, the HST filter combination performs almost as well as HST+JWST/NIRCam in classification accuracy with type resolution (Figure \ref{f:splat:all:DeltaT-metric}). In effect, it appears that kNN classification can be almost as accurate if the HST NIR imaging is sufficiently deep enough to identify the relevant stars. This will be an inherently smaller volume than JWST/NIRCam can reach.

\begin{table}
    \centering
    \begin{tabular}{l | l l l l l l l}
stars & M4 & M8 & L2 & L6 & T0 & T4 & T8 \\
\hline
518 & 491 & 9 & 0 & 2 & 0 & 1 & 15 \\
\hline
    \end{tabular}
    \caption{The numbers of stellar objects identified in the NIRCam fields and the the respective classifications by the KNN. Both M4 and T8 may include types that are hotter and more massive or colder respectively (e.g. K-start in the M0 or Y3-dwarfs in the T8 category).}
    \label{t:ceers:startype}
\end{table}

\section{Results}
\label{s:results}

We will now examine the use of the kNN classification on the NIRCam observations of CEERS described in \S \ref{s:data}.

\subsection{Application to CEERS}

We now apply the kNN classification on the CEERS photometry of unresolved sources identified in Figure \ref{f:ceers:mAB-r50}. Figure \ref{f:ceers:field} shows the distribution of unresolved sources in the ten CEERS fields. We identified a total of 518 stars with a classification ($P_{kNN} > 0.5$) with the majority of probabilities at 0.8 (Table \ref{t:ceers:startype}). These classifications are based on the colours using the F115W, F150W, F200W, and F277W filters. We can now verify these kNN type predictions with the red NIRCam colours.

% \begin{figure}
%     \centering
%     \includegraphics[width=0.5\textwidth]{Figures/Ydwarf.png}
%     \caption{An example of a Y-dwarf from \protect\cite{Leggett21} with some of the absorbing molecule species marked and the CEERS NIRCam filters used annotated at the bottom.  }
%     \label{f:ydwarf}
% \end{figure}

\subsection{NIRCam Red colours}

\begin{figure*}
    \centering
    \includegraphics[width=0.49\textwidth]{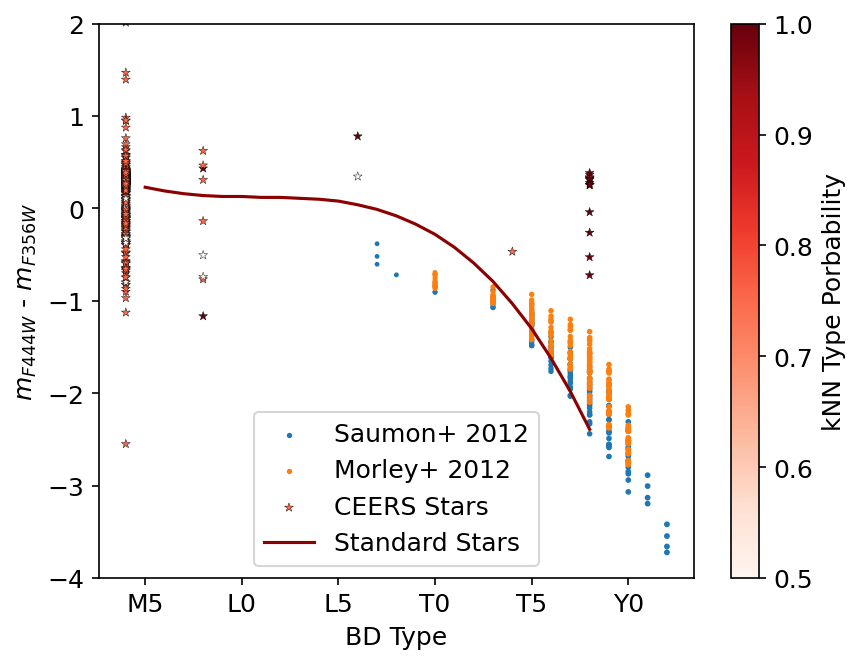}
    \includegraphics[width=0.49\textwidth]{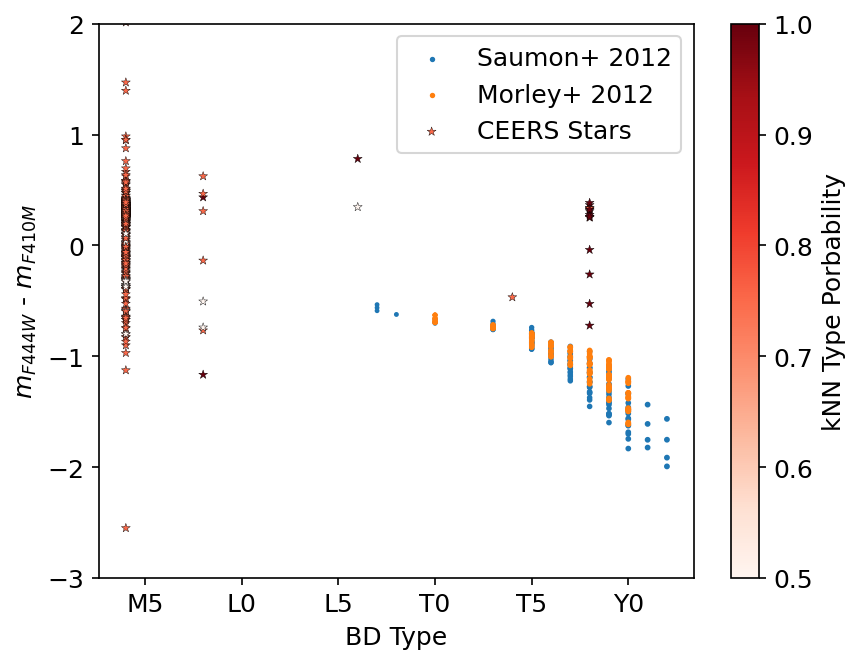}
    \caption{The kNN assigned brown dwarf type against the F444W-F356W and F444W-F410M colours. The dark red line is the W1-W2 WISE colour of standard stars as a function of type.}
    \label{f:ceers:models:knntype-red-colors}
\end{figure*}

Figure \ref{f:ceers:models:knntype-red-colors} shows the kNN type versus the F444W-F356W and F444W-F410M colours. These colours are not included in the SPLAT derived filters as these wavelengths are outside the SpeX spectrograph bandwidth. However, WISE W1-W2 colours are available for the standard stars for each type from \cite{Pecaut13} and these are close stand-ins for the F444W-F356W colour. These colours can be computed for the models of \cite{Saumon12} and \cite{Morley12} L/T/Y atmospheric models as well. The standard stars agree well with these atmospheric models, \rev{specifically those of \cite{Saumon12}. We note that there is almost a magnitude of variance in colour in the models of \cite{Morley12}, reflecting a wider range of parameters.}

The CEERS kNN identified late brown dwarf types ($T>2.5$, T5 dwarf or later) appear to be bluer in the NIRCam reddest filters, the ones not used in the kNN classification. These are some 7 high-confidence \rev{brown dwarfs ($P(T)>80$\%)} that do not line up in the redward colour space. 
\rev{The discrepancy is as much as a magnitude in F444W-F356W colour.}
The remaining late-M and late-T types are much more consistent with the NIRCam long filters. 

Figure \ref{f:ydwarf} gives a hint to the discrepancy between the classification based on the short wavelength filters and the colours observed in the long NIRCam filters. The F410M and F444W both cover a wavelength range where a Sulfur and a Phosphorus molecule is influencing absorption \citep{Leggett21}. If these are T/Y dwarfs with relatively more of such compounds, perhaps it could skew the reddest NIRCam colours \revtwo{towards redder values}?

\rev{The \cite{Saumon12} and \cite{Morley12} models are for Solar metallicity brown dwarfs only. This explains why they align with the standards for brown dwarfs and hints at lower metallicity as a possible explanation for the red colours of these late-type brown dwarfs. Future infrared spectroscopy of these could confirm this.}

\begin{figure*}
    \centering
    \includegraphics[width=0.49\textwidth]{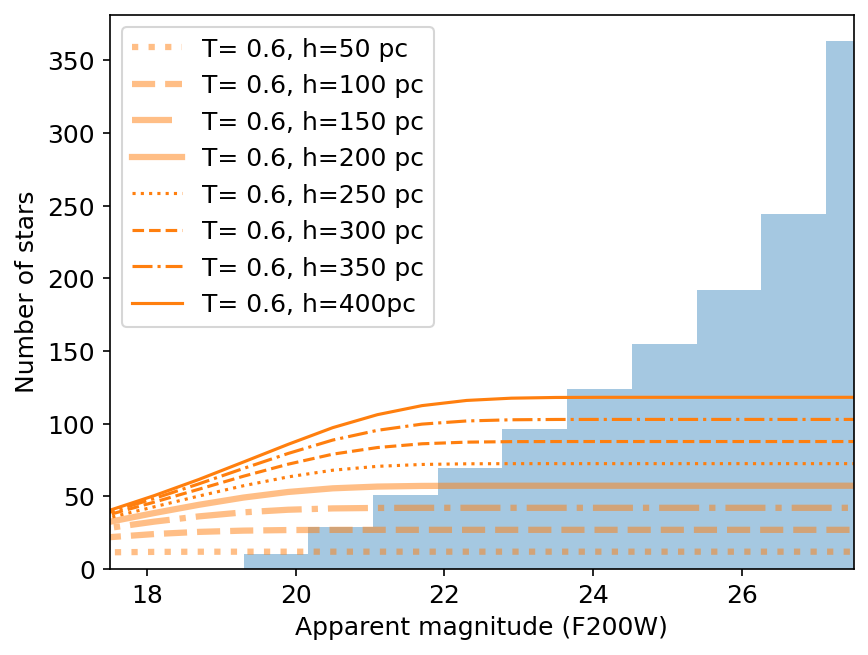}
    \includegraphics[width=0.49\textwidth]{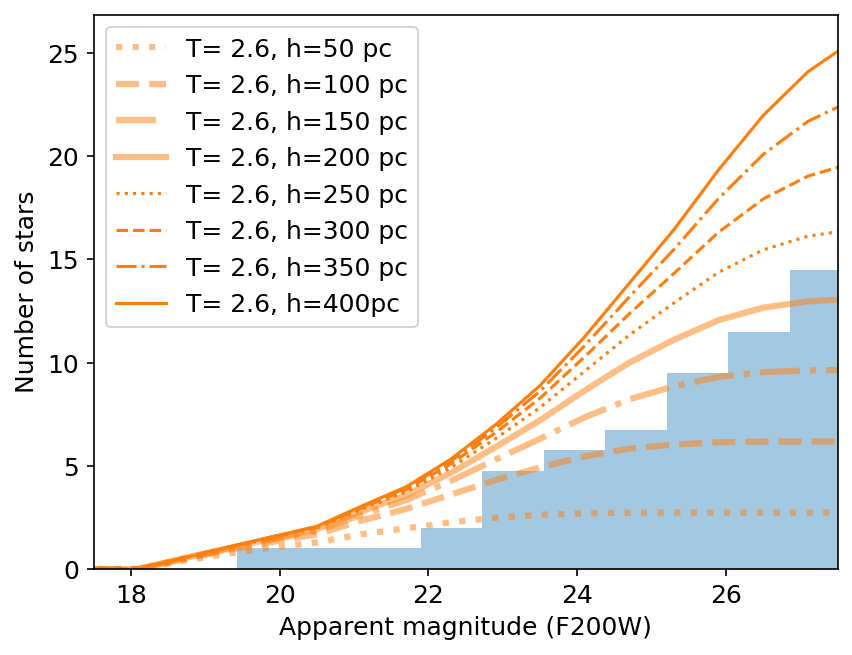}
    \caption{The cumulative histogram of apparent magnitude in F200W for the kNN identified types T=M6$\pm$2 and T=T6$\pm$2 and the expected distribution assuming this type and different Galactic vertical scale-heights.}
    \label{f:ceers:chist:scaleheight}
\end{figure*}

\subsection{\rev{NIRSpec prism observations}}
% \subsection{\rev{NIRSpec prism confirmation}}

\begin{figure*}
    \centering
    \includegraphics[width=0.49\textwidth]{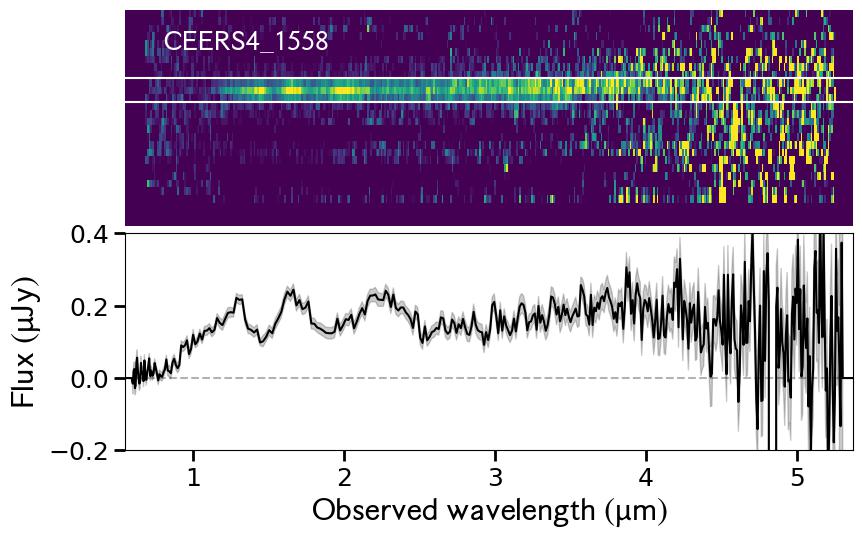}
    \includegraphics[width=0.46\textwidth]{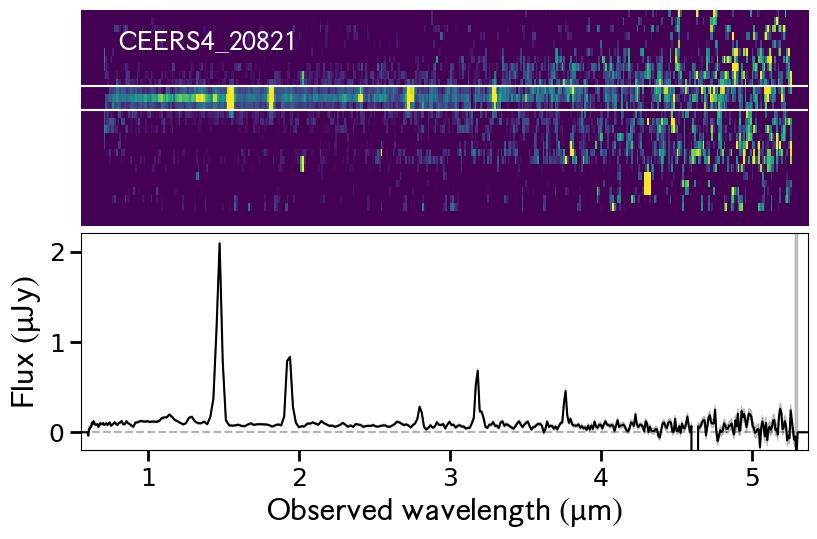}
    \caption{NIRSpec prism spectra of two of the unresolved objects identified by the kNN classification method. The 2D spectra are shown above, with the extraction window marked with a white border. The 1D extracted spectra are shown below. \textit{Left:}   CEERS NIRSpec MSA ID \#1558 shows broad absorption bands at 1.5, 2 and 2.5 microns typical of later type brown dwarfs (see Figure \ref{f:ydwarf}). 
    \textit{Right:} CEERS NIRSpec MSA ID \#20821 is a galaxy at $z = 1.94$ with strong emission lines that perturb the NIRCam photometry, leading to misclassification as a brown dwarf.} 
    % The left panel shows a T-dwarf type objects and the right panel shows an emission line galaxy whose broad colours can be mistaken for a brown dwarf (50\% kNN confidence). }
    \label{f:ceers:nircam-prism}
\end{figure*}

\rev{Two of the kNN brown dwarf candidates were ob0served with CEERS NIRSpec Multi-Shutter Assembly (MSA) spectroscopy (Arrabal Haro et al., \textit{in prep.}) using the low-resolution ($R = 30$--300) prism disperser (Figure \ref{f:ceers:nircam-prism}).  These targets were not originally selected to be stars, but as candidates for distant extragalactic objects.}

% CEERS 53250
\rev{CEERS object \revtwo{53250} (NIRSpec MSA ID 1558) was originally selected as a NIRSpec target from HST CANDELS data as candidate for a galaxy with photometric redshift $z > 6$. Fitting galaxy spectral energy distribution models to the NIRCam and HST photometry yields a photometric redshift sharply peaked at $z\sim 8$ due to the flux discontinuity at $\sim$1$\mu$m.   However, this photometric redshift goodness-of-fit is very poor ($\chi^2 = 240$), and there is a very significant detection in the HST F814W filter, which is inconsistent with a Ly$\alpha$ break at $z \sim 8$.  The stellar kNN classification for is T=0.4 (M4) and T=0.8 (M8) with 50\% probability ($P(T=0.4)=50$\% and $P(T=0.8)=50$\% ). 
The spectrum shows the characteristic absorption features at 1.5 and 2.5 microns typical of a later L- or early T-type (e.g., L4 t0 T2) ultracool dwarf, based on a visual inspection. This is somewhat inconsistent with the confusion matrix of the kNNs (Figure \ref{f:CM:sttype:broad}) where later types are more commonly mis-classified as early-type brown dwarfs.  }

% CEERS id=22148
\rev{CEERS object 22148, with apparent magnitude $m_{F200W} = 26.09$AB and half-light radius $r_{eff} = 0\farcs08$, sits just below the stellar locus identified in Figure \ref{f:ceers:mAB-r50}. Although it was not included in the ``stellar'' object catalog in Table \ref{t:ceers:stars}, we investigated it because it was observed with NIRSpec. The kNN classification for this object is M4/8 with 75/25\% probability ($P(T=0.4)=75$\% and $P(T=0.8)=25$\%). 
It was selected as a NIRSpec target (NIRSpec MSA ID 20821) from HST CANDELS data as a candidate galaxy with photometric redshift $z \approx 1.92$. That identification is confirmed by the NIRSpec spectrum, which shows strong emission lines at $z = 1.94$ from the Hydrogen Balmer and Paschen series, [OIII]4960,5008, [SIII]9533, and He~I~10833. 
%It's apparent magnitude is $m_{F200W} = 26.09$AB and half-light radius is $r_{eff} = 0\farcs08$, sitting just below the stellar locus identified in Figure \ref{f:ceers:mAB-r50}. Not included in the ``stellar'' object catalog in Table \ref{t:ceers:stars}, we investigated it because of the NIRSpec spectrum associated with it. 
%  the kNN classification for this object is M4/8 with 75/25\% probability ($P(T=0.4)=75$\% and $P(T=0.8)=25$\%). 
The emission lines contribute significantly to fluxes measured through the NIRCam filters, perturbing the colors and leading to the photometric misclassification as a cool star. This object serves as an important reminder that individual kNN identified objects are primarily \textit{candidates} until spectroscopically confirmed, and that careful selection of unresolved objects can aid in the precision of the selected sample. Fortunately, a probabilistic Milky Way model can cope with a certain amount of contamination. For example, the MCMC model from \cite{van-Vledder16} accounted for 40\% misclassification. }

%  \rev{Figure \ref{f:ceers:nircam-prism} shows one of the CEERS NIRSpec prism spectra of the candidate brown dwarfs identified with the kNN method in the CEERS field. The kNN classification is T=0.4 (M4) with 50\% probability ($P(T)=50$\%). This spectrum of CEERS object \revtwo{53250} (spec ID 1558) shows the characteristic absorption features at 1.5 and 5 microns typical of an early T-type (e.g T2) ultracool dwarf. And this is consistent with the confusion matrix of the kNNs (Figure \ref{f:CM:sttype:broad}) where earlier types are mis-classified as late-type brown dwarfs. The misclassifications occur more the other way.}

% \rev{A second object appears to be closer to an emission line object which is an important reminder that individual kNN identified objects are candidate objects until spectroscopically confirmed, similar to procedure for the high redshift objects or dusty dwarf galaxies at $z\sim2$. However, a Milky Way model such as the MCMC one in \cite{van-Vledder16} can deal with a certain amount of contamination. }

\subsection{Scale-height of Brown Dwarfs}

\rev{To model the scale-height of the Galactic disk along the one line-of-sight of the CEERS field, we use the cumulative count of brown dwarfs as a function of apparent magnitude \citep[similar to][]{Pirzkal05,Pirzkal09}. Assuming a brown dwarf type and the local density \revtwo{within 20 parsec} from \cite{Reid08} and \cite{Kirkpatrick21} and a scale-height, one can compute the numbers of brown dwarfs along the line-of-sight. This works best with a single brown dwarf type, a cleanly selected sample, and only the thin disk of the Milky Way considered. }

Figure \ref{f:ceers:chist:scaleheight} shows two cumulative histograms of the brown dwarf candidates identified in CEERS. The early types (T=0.6, i.e. M6) show an over-density compared to the thin disk. \rev{At the lower apparent luminosities, there is a clear over-density that cannot be explained by a Galactic thin disk alone.}
This is similar to what \cite{Holwerda14a} and \cite{van-Vledder16} found for M-dwarfs in the HST pure-parallel fields. The cumulative histogram is consistent with a thin disk with a scale-height of $z_0=250-300$pc, similar to the scale-heights found before \citep{Ryan05,Ryan11,Pirzkal05,Pirzkal09,Holwerda14,van-Vledder16,Aganze22b}. 
A second Galactic structural component is needed to explain the numbers of M-dwarfs in deep JWST/NIRCam fields. 

The second panel of Figure \ref{f:ceers:chist:scaleheight} shows the distribution of late dwarfs (T=2.6, i.e. T6$\pm$2). Their distribution is mostly consistent with a thin disk alone as they are intrinsically much fainter ($M_J= 17.06$AB). \rev{A second component is not fully evident from the counts in CEERS.} The cumulative histogram is consistent with a thin disk with a scale-height of $z_0=150-200$pc, similar to the scale-height of 175pc found before \citep{Ryan11,Aganze22b}. 

There is the matter of types in-between the early and late brown dwarfs apparently missing in this field. There are two L-dwarfs and one early T-dwarf in this field with reasonably high confidences. This is not yet enough statistics to infer a scale-height with much confidence. To infer scale-heights accurately, one will need many more lines-of-sight though the Milky Way disk.

\section{Discussion}

JWST opens up a new parameter space in the near-infrared,  thanks to both more collecting area and much improved efficiency compared to other NIR observatories. CEERS is meant to demonstrate this potential, especially in extra-Galactic science. However, the depth and colour resolution that benefits extra-galactic science can benefit science examining our own Milky Way's stellar population just as much. 

Figure \ref{f:ceers:mAB-r50} illustrates the first improvement: the identification of stellar objects in drizzled NIRCam images increases from HST ($\sim25$) to $\sim$28AB mag using otherwise identical techniques. This alone opens a wide volume of the Galaxy for low-mass stellar objects. 

The much greater filter suite available onboard NIRCam is a second capability that expands our discovery space over HST/WFC3. The bluest NIRCam filters combine into a similar feature space (Figure \ref{f:splat:IDE:jwst-only} compared to Figure \ref{f:splat:IDE:hst-only}) but the addition of longer wavelength information with NIRCam immediately allows us to check kNN classifications. 
For these colors atmosphere models can stand in for the SPLAT observations (see Appendix A) but performance is not quite as reliable as SPLAT. This can be attributed to the much larger training sample in SPLAT ($\sim$1200) over these models ($\sim$300) which captures the range of characteristics inherent in the population. The models however extend to later types of brown dwarfs and cover longer wavelengths, opening up the possibility to train on models and then use the full NIRCam wavelength range to classify stellar objects. 

The late-type candidates for example appear to have too blue colours in the reddest NIRCam filters (Figure \ref{f:ceers:models:knntype-red-colors}). This could be mis-identification or an issue with calibration of the kNN. The effect is quite pronounced and could speculatively be explained by more sulfur and phosphorus compounds in these substellar objects (Figure \ref{f:ydwarf}). 

\begin{figure*}
    \centering
    \includegraphics[width=0.49\textwidth]{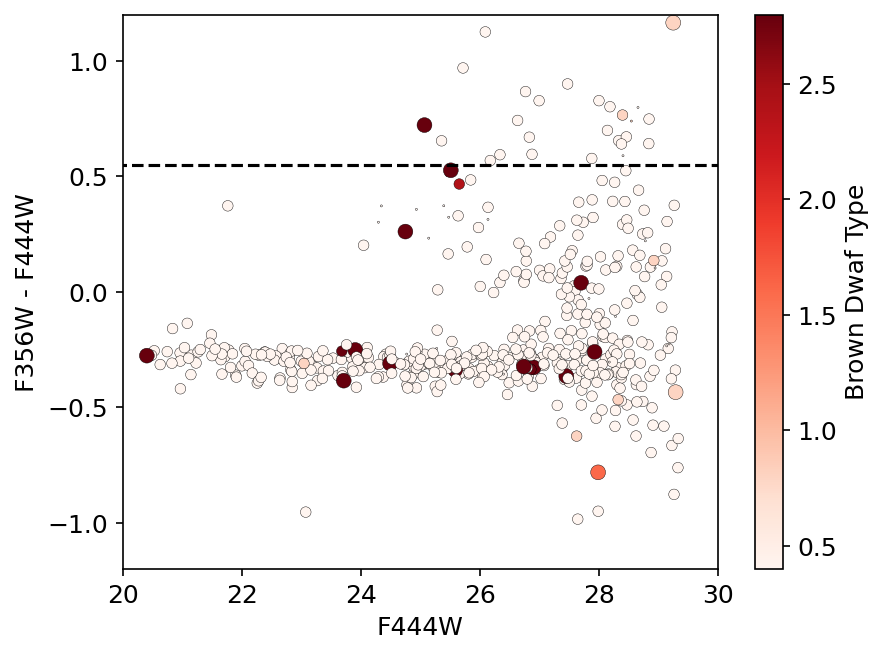}
    \includegraphics[width=0.49\textwidth]{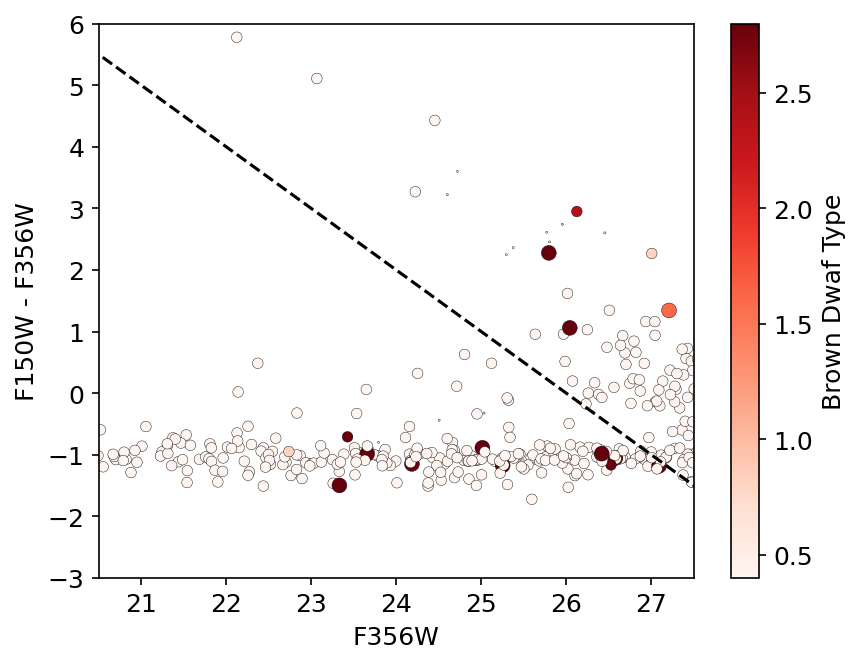}
    \caption{A comparison to Figure 4 of \protect\cite{Bisigello23} and Figure 1 from \protect\cite{Perez-Gonzalez23a} for the kNN classified stellar objects. \revtwo{Marker size is indicative of the kNN probability assigned to each object's classification.} The dashed lines are the lower limits for the selection criteria for dusty dwarfs (left) or HST-dark galaxies (right). Both populations can suffer from contamination from brown dwarfs. }
    \label{f:ceers:literature:comp}
\end{figure*}

\rev{
The F444W-F356W colours of the brown dwarfs are very similar to the F200W dropout galaxies identified in \cite{Bisigello23}, their figure 4 (see Figure \ref{f:ceers:literature:comp}) \revtwo{but with F200W and bluer detections}. The accurate identification of brown dwarfs is therefore critical in the removal of potential interlopers in the identification of new (low redshift) dwarf galaxy populations ($z<2$). Similarly, some kNN identified late-type ultracool dwarfs reside in the HST-dark section of the colour space identified in \cite{Perez-Gonzalez23a}. These authors took care to select these samples using other colour cuts and SED fits. Figure \ref{f:ceers:literature:comp} illustrates the possibility that brown dwarfs could be included without such precautions. In such cases, a high probability assigned by the kNN classifier that the unresolved object is a brown dwarf may be a good reason to remove an object from a selection. }

% Maybe I missed it but I did not see a plot involving F356W-F150W, which is what I used in my selection. Apart from that, there were full spectral fitting analysis and I don't think there can be any confusion. But....
% The samples that can be contaminated by these BD are the very high-z ones. There we only have 2-3 colors, and we normally impose something like F277W-F200W or F200W-F150W redder than 1.0 or 1.5. There you can have a DB, you have 1-2 examples, I think.

\begin{figure*}
    \centering
    \includegraphics[width=0.49\textwidth]{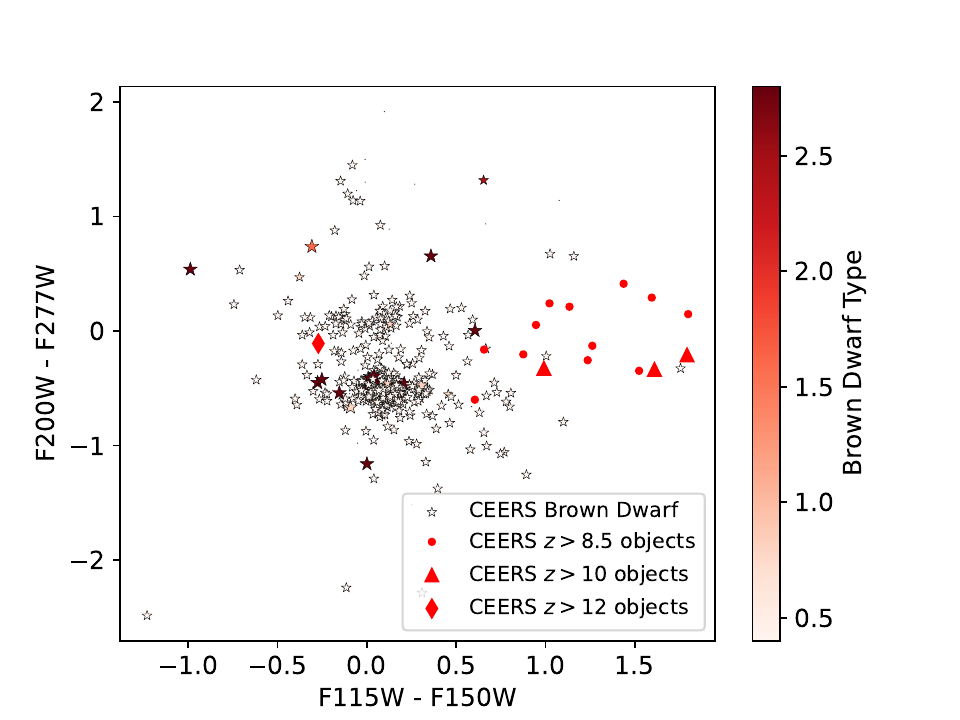}
    \includegraphics[width=0.49\textwidth]{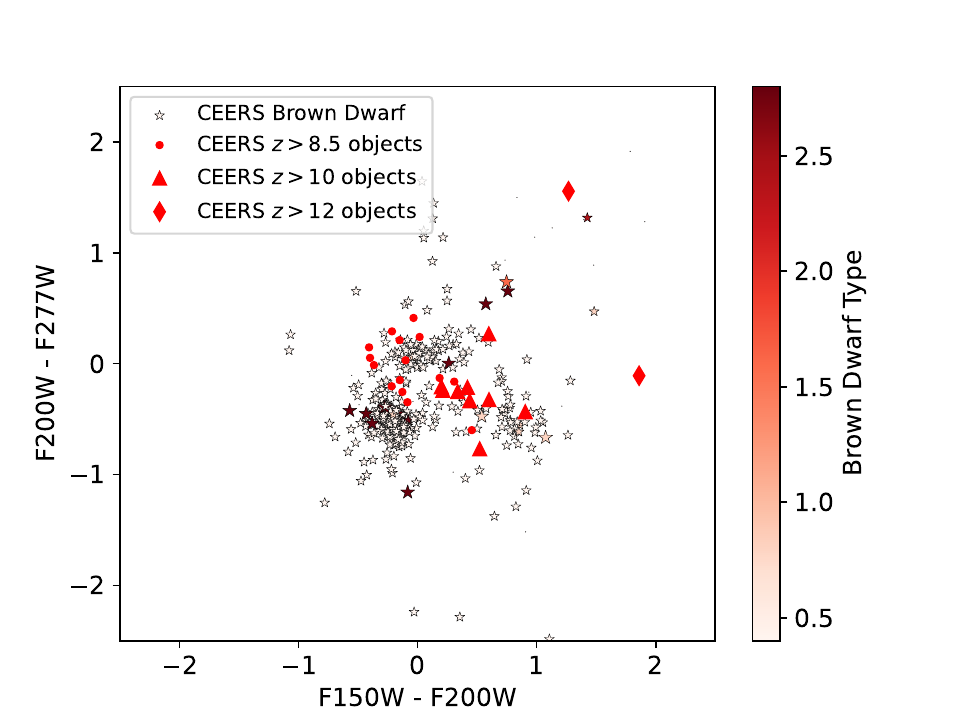}
    \caption{The F115W-F150W and the F150W-F200W vs F200W-F277W colours of the brown dwarf candidates identified in this paper and the same colours of the \protect\cite{Finkelstein23} selection of high ($z>8$) redshift galaxies. }
    \label{f:ceers:highz:comp}
\end{figure*}

\revtwo{ 
A major consideration in the successful identification of brown dwarfs in extra-galactic JWST data is to remove them as contaminant in searches for high-redshift sources ($z>8$)\footnote{The definition of the term ``high-redshift" has changed in the era of JWST but this is the one we adopt here.} 
Figure \ref{f:ceers:highz:comp} shows the bluer NIRCam colours of the brown dwarfs identified with kNN and the high-redshift sources from \cite{Finkelstein23}. The $z>12$ sources are dropouts in the F115W and therefore not shown in the bluest colour F115W-F150W. Both the high-redshift sample and the brown dwarf kNN selection do not rely on hard colour criteria but instead on good SED fits to a template and redshift or a close set of kNN templates in colour space.}

\revtwo{Figure \ref{f:ceers:highz:comp} illustrates that a colour-cut in F115W-F150W can separate the two popylatins well but that the redder NIRCam colours are very similar for brown dwarfs and high-redshift galaxies alike. 
}

\rev{
Photometric identification of brown dwarfs in JWST data is ongoing \citep{Nonino23, Beiler23, Wang23e}. Like all photometric identifications, the kNN classifications here would benefit from spectroscopic confirmation using NIRSpec. Figure \ref{f:ceers:nircam-prism} shows a first such spectroscopic confirmation with a low-resolution NIRSpec prism spectrum. This would be the second ultracool brown dwarf spectroscopically confirmed with JWST after \cite{Beiler23}.}

\rev{
\cite{Wang23e} identify a T dwarf in the public CEERS data. This object is not in our kNN classified sample as it too faint to be included ($m_{F200W}=28.45$AB, see Figure \ref{f:ceers:mAB-r50}). However, a kNN classification trained on just the blue JWST/NIRCam filters (F115W, F150W, F200W, and F277W) predicts a L7 dwarf ($T_{eff} = 1530$K, $P_{kNN}(Type)=100$\%), a higher effective temperature than \citep[$T_{eff}=1300$, T0][]as classified by {Wang23e}. The reported F444W-F356W colour (-0.14) is consistent with the kNN classification as well as the \cite{Wang23e} one (Figure \ref{f:ceers:models:knntype-red-colors}). We can note here that L7 and T0 are less than 4 subtypes apart. }

\rev{The difference between kNN classification and spectroscopic classification for individual sources highlights the statistical nature of the kNN classifications. It shows that kNN classifications are good enough for a population of sources to be used in a statistical manner to infer the shape of the Milky Way. An MCMC model such as employed by \cite{van-Vledder16} has the benefit of explicitly dealing with a fraction of the data as bad as a fit parameter. }

% \cite{Nonino23} - GLASS-JWST grism 
% Bisigello+23 Beiler+ 23 (see their Appendix C); and Wang, Goto+23), b

% Miles+23 observed a known brown dwarf and obtained lovely NIRSpec and MIRI MRS spectra. Roellig+23 (AAS presentation) observed a known T dwarf rotator and obtained time-resolved NIRSpec and MIRI spectra.

The inferred vertical profiles of M and T-dwarfs is 300-350pc for the $M6\pm2$ dwarf candidates in the CEERS field with a clear second component starting at $m_{F200W}\sim 24$ in Figure \ref{f:ceers:chist:scaleheight}. This is consistent with what \cite{van-Vledder16} found for HST/WFC3 pure-parallel fields (300pc). The scale-height of the candidate T-dwarfs is more consistent with 150-200pc. This is remarkably consistent with what \cite{Aganze22b} found from HST grism observations (175pc). 

The photometric identification of M-L-T-Y dwarf stars in deep extragalactic observations such as CEERS has a success rate (85\% or better) with existing photometry. It seems only a matter of time before enough sightlines through the Milky Way will be available for an accurate assessment of their distribution and total numbers. 

% Wang23e found a T-dwarf in CEERS. Discussd

\section{Conclusions}

We have applied kNN identificataion of brown dwarf types on the unresolved sources in the stellar locus of CEERS photometry with the aim of identifying this Galactic population 

Broad filters onboard JWST/NIRCam are sufficient to identify brown dwarfs photometrically to within 2 subtypes using k-Nearest Neighbour classification using the four nearest neighbours. We only use the four bluest filters for this (F115W, F150W, F200W and F277W). The SPLAT library is optimal as a training set with sufficient size and coverage of type. Conceivably atmospheric models of M-L-T-Y dwarfs may replace these as the training set.  

% We can identify these low-mass and substellar objects out to XX kpc with JWST/NIRCam imaging.

The late-type brown dwarfs identified in CEERS are bluer in the NIRCam red filters than expected on the kNN classification based on NIR blue filters. This remains unexplained by standard stars or atmospheric models. 

The scale height of the MW disk is consistent with the 350 pc for type M-types, similar to what \cite{van-Vledder16} found. The vertical distribution of later T-types is consistent with a scale-height of 150-200pc, in line with the value from \cite{Aganze22b}.

\section*{Acknowledgements}

The author would like to thank the CEERS team for their hard work making the final exquisite  data available for all to use. 

%%%%%%%%%%%%%%%%%%%%%%%%%%%%%%%%%%%%%%%%%%%%%%%%%%
\section*{Data Availability}

CEERS is committed to data-releases. This work is based on the data released in the first public data-release. See the CEERS website for details: \url{https://ceers.github.io/releases.html}.

%%%%%%%%%%%%%%%%%%%% REFERENCES %%%%%%%%%%%%%%%%%%

% The best way to enter references is to use BibTeX:

%\bibliographystyle{mnras}
%\bibliography{Bibliography} % if your bibtex file is called example.bib

% Alternatively you could enter them by hand, like this:
% This method is tedious and prone to error if you have lots of references
%\begin{thebibliography}{99}
%\bibitem[\protect\citeauthoryear{Author}{2012}]{Author2012}
%Author A.~N., 2013, Journal of Improbable Astronomy, 1, 1
%\bibitem[\protect\citeauthoryear{Others}{2013}]{Others2013}
%Others S., 2012, Journal of Interesting Stuff, 17, 198
%\end{thebibliography}

%%%%%%%%%%%%%%%%%%%%%%%%%%%%%%%%%%%%%%%%%%%%%%%%%%
\clearpage
\newpage
%%%%%%%%%%%%%%%%% APPENDICES %%%%%%%%%%%%%%%%%%%%%

\appendix

\section{Models}

\begin{figure*}
    \centering
    \includegraphics[width=\textwidth]{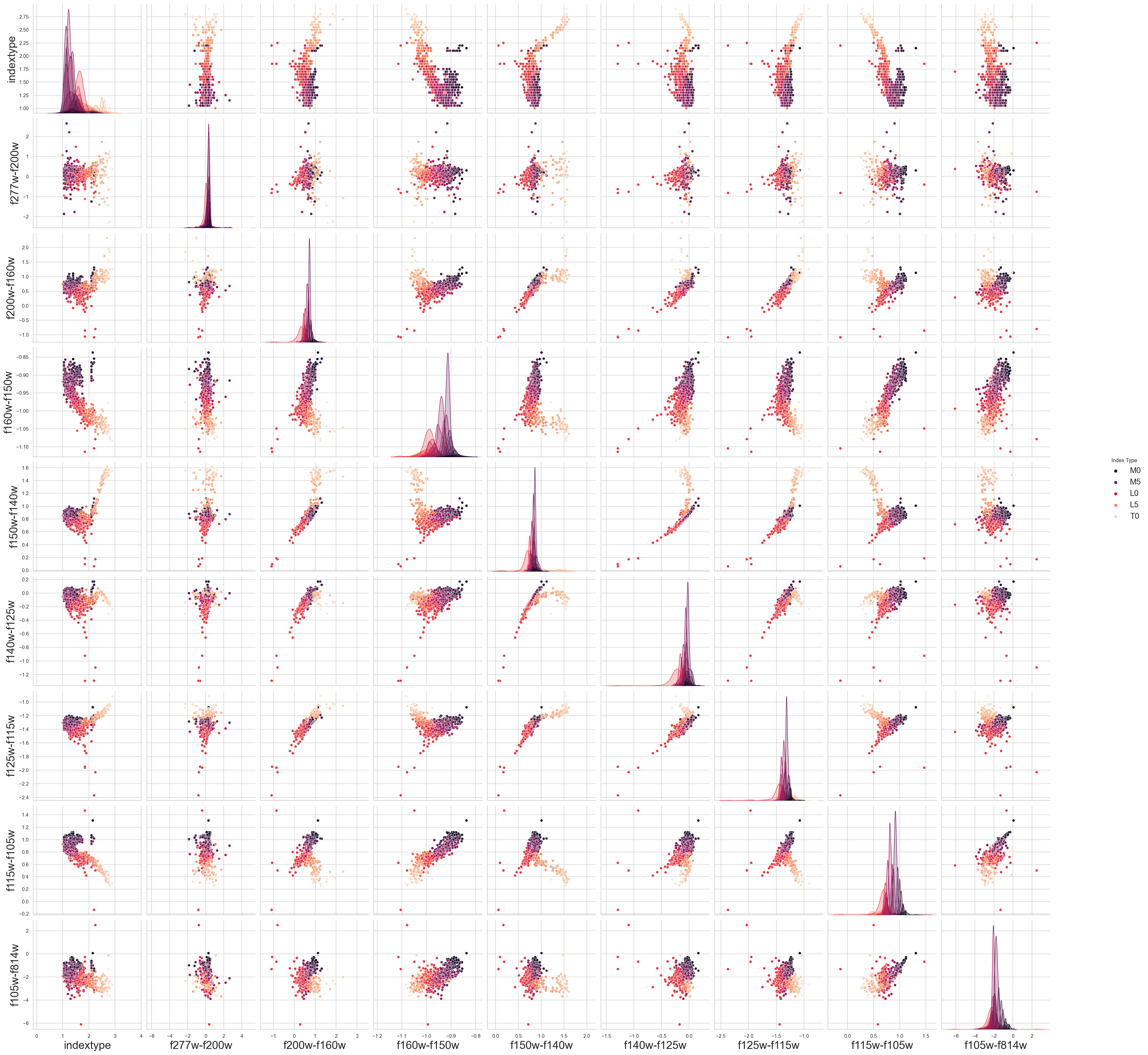}
    \caption{The colour-colour feature space from the models of \protect\cite{Saumon12} and \protect\cite{Morley12}. The type space is a few late L-type, most T-types and early Y-type. The full feature space does include all CEERS filter colour combinations. }
    \label{f:models:feature}
\end{figure*}

We test the kNN classifier based on the models from \protect\cite{Saumon12} and \protect\cite{Morley12}. In total 337 different models are available. Figure \ref{f:models:feature} shows the colour feature space. Brown dwarf type was assigned to each model using the table from \cite{Pecaut13} matching model temperature to a brown dwarf type with similar temperature. This is not as big a training set as the SPLAT catalogue.

\begin{figure}
    \centering
    \includegraphics[width=0.5\textwidth]{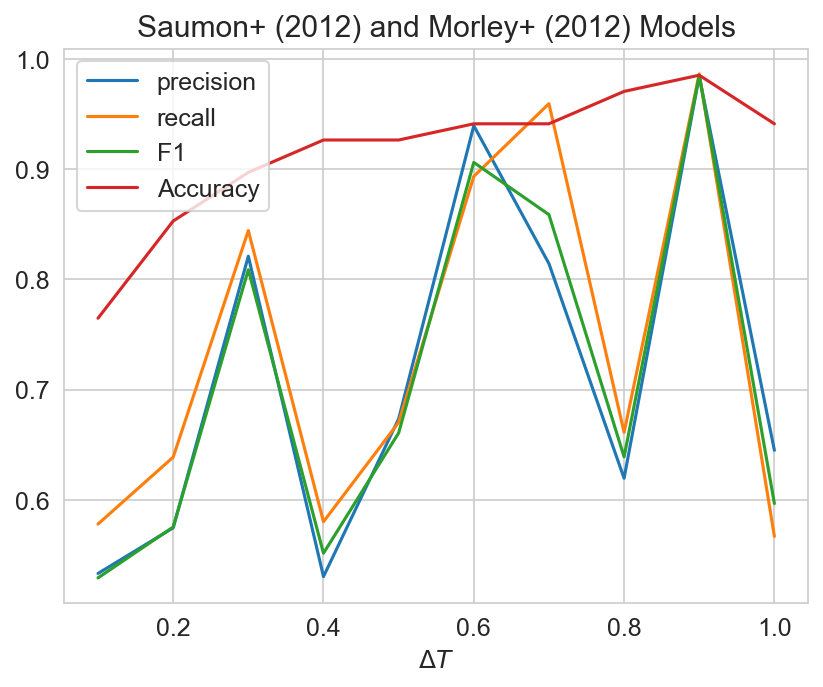}
    \caption{The performance of the kNN using the colours based on the models of \protect\cite{Saumon12} and \protect\cite{Morley12}. The higher dimensionality but lower type coverage results in some erratic behavior. }
    \label{f:models:performance}
\end{figure}

\begin{figure}
    \centering
    \includegraphics[width=0.5\textwidth]{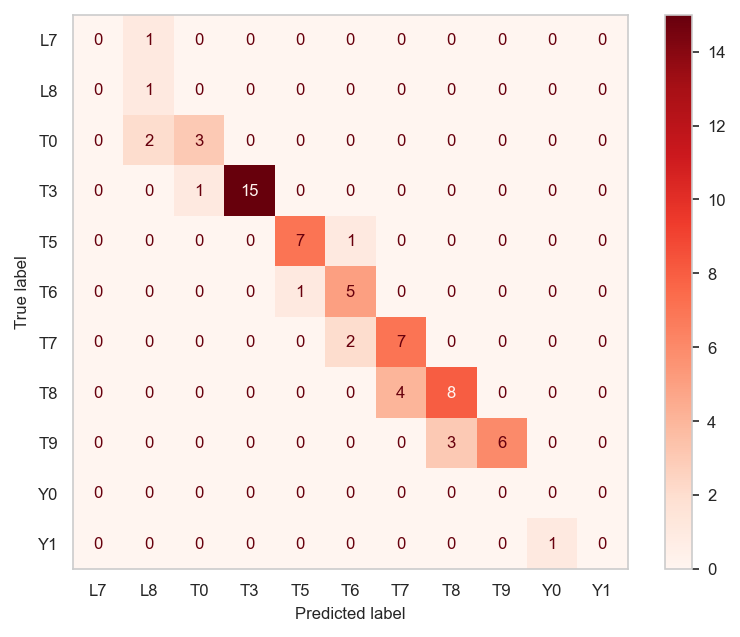}
    \caption{The confusion matrix of the kNN model based on the colours based on the models of \protect\cite{Saumon12} and \protect\cite{Morley12}. Compared to the SPLAT confusion matrix in Figure \ref{f:CM:sttype:full}. }
    \label{f:models:CM}
\end{figure}

Figure \ref{f:models:CM} shows the confusion matrix of the models. Like the SPLAT catalogue, there is a tendency for the kNN to classify brown dwarfs slightly later. Coverage of type is not as uniform as SPLAT and this hinders interpolation with kNN some. The performance as a function of type resolution is shown in Figure \ref{f:models:performance} and this lack of coverage is evident in an erratic increase in most metrics.

There are two major benefits of these models over the SPLAT catalog: first the feature space includes the red NIRCam filters as well. One can use all the NIRCam photometry to classify. Secondly, the models extend to much later types than SPLAT. Future suites of models \revtwo{(e.g. improved line lists, disequilibrium chemistry, improved treatment of clouds etc.)}, once verified with spectroscopy, could well serve as the training set for future kNN classifications. 
 
\clearpage

\begin{table*}
    \centering
\begin{tabular}{l l l l l l l l l l l l l l l l l l}
ID & T & P(T) & RA & DEC & $m_{606}$ & $m_{814}$ & $m_{105}$ & $m_{125}$ & $m_{140}$ & $m_{160}$ & $m_{115}$ & $m_{150}$ & $m_{200}$ & $m_{277}$ & $m_{356}$ & $m_{410}$ & $m_{444}$ \\     
\hline
\hline
230 &  0.4 & 0.75 & 215.005725 & 53.019548 & 28.2 & 27.9 &  \dots &  \dots & 26.6 &  \dots & 27.0 & 26.9 & 27.3 & 27.6 & 28.0 & 28.4 & 28.6 \\
442 &  0.4 & 0.75 & 214.984572 & 53.003732 & 28.0 & 28.0 &  \dots &  \dots &  \dots &  \dots & 27.9 & 28.0 & 27.9 & 27.8 & 27.4 & 28.1 & 28.0 \\
471 &  0.4 & 0.75 & 215.008698 & 53.019115 & 22.4 & 20.5 &  \dots & 19.6 & 19.4 & 19.3 & 19.5 & 19.3 & 19.4 & 20.1 & 20.3 & 20.3 & 20.5 \\
1056 &  0.4 & 0.75 & 214.979593 & 52.996941 & 28.0 & 27.5 &  \dots & 27.3 & 28.1 & 27.9 & 27.7 & 27.5 & 26.8 & 27.2 & 28.0 & 27.9 & 27.7 \\
1273 &  0.4 & 0.75 & 214.988092 & 53.002487 & 27.1 & 26.8 &  \dots &  \dots & 27.2 &  \dots & 26.8 & 26.9 & 26.9 & 26.8 & 27.0 & 27.3 & 27.3 \\
1411 &  0.4 & 0.75 & 214.989956 & 53.002400 & 26.2 & 24.9 &  \dots & 23.7 & 23.5 & 23.6 & 23.4 & 23.3 & 23.5 & 24.0 & 24.3 & 24.4 & 24.6 \\
1576 &  0.4 & 0.75 & 214.960103 & 52.980446 & 26.9 & 25.2 & \dots & 24.7 & 24.7 & 24.7 & 24.5 & 24.5 & 24.7 & 25.1 & 25.5 & 25.7 & 25.8 \\
\hline
    \end{tabular}
    \caption{-- \textit{continued}}
\end{table*}

\setcounter{table}{0}

\begin{table*}
    \centering
    \begin{tabular}{l l l l l l l l l l l l l l l l l l}
ID & T & P(T) & RA & DEC & $m_{606}$ & $m_{814}$ & $m_{105}$ & $m_{125}$ & $m_{140}$ & $m_{160}$ & $m_{115}$ & $m_{150}$ & $m_{200}$ & $m_{277}$ & $m_{356}$ & $m_{410}$ & $m_{444}$ \\     
\hline
\hline
14111 &  0.4 & 0.75 & 214.866930 & 52.904939 & 26.2 & 25.3 & \dots & 24.6 & 24.7 & 24.6 & 24.6 & 24.5 & 24.8 & 25.2 & 25.6 & 25.7 & 25.9 \\
14129 &  0.4 & 0.75 & 214.910007 & 52.935116 & 28.1 & 27.7 & \dots & 27.6 & 28.1 & 27.5 & 27.6 & 27.6 & 27.5 & 27.5 & 27.9 & 28.0 & 28.1 \\
14185 &  0.4 & 0.75 & 214.876678 & 52.911429 & \dots & 28.9 & \dots & 27.3 & 28.0 & 27.1 & 27.4 & 27.6 & 27.7 & 27.9 & 27.8 & 27.7 & 28.1 \\
14865 &  0.4 & 0.75 & 214.880989 & 52.910540 & 30.3 & 29.0 & \dots & 28.6 & 28.0 & 27.9 & 28.4 & 28.2 & 27.7 & 27.4 & 27.4 & 27.5 & 27.4 \\
14968 &  0.4 & 0.75 & 214.910169 & 52.930537 & 24.4 & 23.4 & \dots & 22.9 & 22.9 & 22.9 & 22.9 & 22.8 & 22.9 & 23.5 & 23.8 & 23.9 & 24.1 \\
15645 &  0.4 & 0.75 & 214.930282 & 52.941671 & 27.8 & 26.5 & \dots & 25.8 & 26.3 & 25.9 & 25.9 & 25.8 & 26.1 & 26.5 & 26.9 & 27.1 & 27.2 \\
16420 &  0.4 & 0.75 & 214.866867 & 52.891148 & 28.5 & 28.2 & \dots & 27.6 & 27.8 & 27.8 & 27.8 & 27.7 & 27.6 & 27.9 & 28.4 & 28.6 & 28.5 \\
\hline
    \end{tabular}
    \caption{-- \textit{continued}}
\end{table*}

\setcounter{table}{0}

\begin{table*}
    \centering
    \begin{tabular}{l l l l l l l l l l l l l l l l l l}
ID & T & P(T) & RA & DEC & $m_{606}$ & $m_{814}$ & $m_{105}$ & $m_{125}$ & $m_{140}$ & $m_{160}$ & $m_{115}$ & $m_{150}$ & $m_{200}$ & $m_{277}$ & $m_{356}$ & $m_{410}$ & $m_{444}$ \\     
\hline
\hline
27227 &  0.4 & 0.75 & 214.769238 & 52.823309 & 25.6 & 25.1 & \dots & 24.9 & 24.8 & 24.8 & 24.8 & 24.8 & 25.0 & 25.6 & 26.1 & 26.3 & 26.5 \\
27283 &  0.4 & 0.75 & 214.787729 & 52.836066 & 24.6 & 23.5 & \dots & 22.9 & 22.9 & 22.8 & 23.1 & 22.7 & 22.9 & 23.4 & 23.7 & 23.9 & 24.1 \\
27458 &  0.4 & 0.75 & 214.766870 & 52.820418 & 28.2 & 26.8 & \dots & 26.0 & 26.6 & 26.1 & 26.0 & 25.9 & 26.1 & 26.6 & 26.9 & 27.0 & 27.1 \\
27809 &  0.4 & 0.75 & 214.848625 & 52.876818 & 26.9 & 27.0 & \dots & 26.4 & 26.6 & 27.0 & 26.4 & 26.9 & 26.9 & 26.7 & 26.9 & 26.8 & 26.7 \\
27873 &  0.4 & 0.75 & 214.848537 & 52.876806 & 27.5 & 27.5 & \dots & 27.0 & 27.0 & 27.3 & 27.0 & 27.3 & 27.3 & 27.2 & 27.1 & 27.1 & 27.1 \\
28165 &  0.4 & 0.75 & 214.823380 & 52.857420 & 24.0 & 23.6 & \dots & 23.5 & 23.4 & 23.4 & 23.4 & 23.3 & 23.6 & 24.2 & 24.7 & 24.9 & 25.1 \\
28264 &  0.4 & 0.75 & 214.828982 & 52.861133 & 28.0 & 27.8 & \dots & 28.8 & 27.4 & 27.4 & 27.8 & 27.6 & 26.9 & 27.3 & 27.9 & 27.9 & 27.8 \\
\hline
    \end{tabular}
    \caption{-- \textit{continued}}
\end{table*}

\setcounter{table}{0}

\begin{table*}
    \centering
    \begin{tabular}{l l l l l l l l l l l l l l l l l l}
ID & T & P(T) & RA & DEC & $m_{606}$ & $m_{814}$ & $m_{105}$ & $m_{125}$ & $m_{140}$ & $m_{160}$ & $m_{115}$ & $m_{150}$ & $m_{200}$ & $m_{277}$ & $m_{356}$ & $m_{410}$ & $m_{444}$ \\     
\hline
\hline
37455 &  0.4 & 0.75 & 214.780023 & 52.781031 & 31.3 & \dots & \dots & \dots & 28.5 & 29.2 & 27.4 & 27.5 & 27.5 & 29.8 & 24.2 & \dots & 30.4 \\
38393 &  0.4 & 0.50 & 214.806850 & 52.795130 & 28.3 & 28.4 & \dots & 28.0 & \dots & 28.0 & 28.8 & 28.2 & 27.4 & 28.1 & 29.0 & 28.8 & 28.9 \\
38499 &  0.4 & 0.75 & 214.734404 & 52.742211 & 28.3 & 28.4 & \dots & 28.7 & 28.4 & 28.3 & 28.3 & 28.1 & 27.7 & 27.9 & 28.2 & 28.2 & 28.1 \\
38545 &  0.4 & 0.75 & 214.720979 & 52.732671 & 29.3 & 27.9 & \dots & 27.7 & \dots & 27.2 & 28.2 & 27.2 & 27.7 & 28.0 & 28.4 & 28.7 & 28.7 \\
38589 &  0.4 & 0.75 & 214.729152 & 52.738314 & 29.6 & 27.8 & \dots & 27.3 & 27.3 & 27.5 & 27.3 & 27.2 & 27.5 & 27.9 & 28.2 & 28.4 & 28.6 \\
38781 &  0.4 & 0.75 & 214.735560 & 52.751323 & 23.3 & 22.7 & \dots & 22.3 & 22.3 & 22.3 & 22.3 & 22.2 & 22.4 & 22.9 & 23.3 & 23.5 & 23.7 \\
39222 &  0.4 & 0.75 & 214.801223 & 52.786505 & 27.7 & 26.9 & \dots & 26.4 & 26.4 & 26.8 & 26.3 & 26.3 & 26.5 & 27.0 & 27.6 & 27.5 & 27.6 \\
\hline
    \end{tabular}
    \caption{-- \textit{continued}}
\end{table*}

\setcounter{table}{0}

\begin{table*}
    \centering
    \begin{tabular}{l l l l l l l l l l l l l l l l l l}
ID & T & P(T) & RA & DEC & $m_{606}$ & $m_{814}$ & $m_{105}$ & $m_{125}$ & $m_{140}$ & $m_{160}$ & $m_{115}$ & $m_{150}$ & $m_{200}$ & $m_{277}$ & $m_{356}$ & $m_{410}$ & $m_{444}$ \\     
\hline
\hline
51556 &  0.4 & 0.75 & 214.922952 & 52.853371 & \dots & \dots & \dots & \dots & \dots & 28.0 & \dots & \dots & 27.8 & 29.9 & \dots & 28.2 & \dots \\
51848 &  0.4 & 0.75 & 214.795081 & 52.817083 & 26.7 & 23.3 & \dots & 21.7 & 21.7 & 21.7 & 21.5 & 21.5 & 21.6 & 22.1 & 22.1 & 22.2 & 22.4 \\
51940 &  0.4 & 0.75 & 214.852020 & 52.857075 & 20.4 & 19.9 & \dots & 19.6 & 19.5 & 19.5 & 19.8 & 19.6 & 19.8 & 20.5 & 20.9 & 21.0 & 21.2 \\
51984 &  0.4 & 0.75 & 214.855830 & 52.859984 & 29.0 & 29.0 & \dots & 27.8 & 27.9 & 27.3 & 27.8 & 27.2 & 26.8 & 26.7 & 26.5 & 26.5 & 26.4 \\
52031 &  2.8 & 1.00 & 214.857516 & 52.861120 & 28.8 & 27.5 & \dots & 28.2 & 25.6 & 26.4 & 27.1 & 28.1 & 27.5 & 27.0 & 25.8 & 25.2 & 25.1 \\
52223 &  0.4 & 0.75 & 214.854426 & 52.857813 & 20.1 & 19.9 & \dots & 20.0 & 20.1 & 20.1 & 20.0 & 20.1 & 20.5 & 21.3 & 21.5 & 21.7 & 21.9 \\
52259 &  0.4 & 0.75 & 214.793015 & 52.813822 & 29.3 & 28.4 & \dots & 29.7 & \dots & 28.1 & 27.7 & 27.7 & 27.8 & 28.3 & 28.6 & 29.0 & 29.2 \\
\hline
    \end{tabular}
    \caption{-- \textit{continued}}
\end{table*}

\setcounter{table}{0}

\begin{table*}
    \centering
    \begin{tabular}{l l l l l l l l l l l l l l l l l l}
ID & T & P(T) & RA & DEC & $m_{606}$ & $m_{814}$ & $m_{105}$ & $m_{125}$ & $m_{140}$ & $m_{160}$ & $m_{115}$ & $m_{150}$ & $m_{200}$ & $m_{277}$ & $m_{356}$ & $m_{410}$ & $m_{444}$ \\     
\hline
\hline
65296 &  0.4 & 0.75 & 215.122474 & 52.981160 & 27.6 & 27.5 & \dots & 26.9 & \dots & 26.7 & 27.0 & 26.5 & 26.3 & 26.1 & 26.0 & 25.9 & 25.8 \\
65330 &  0.4 & 0.75 & 215.074800 & 52.947440 & 27.3 & 26.2 & 25.9 & 25.9 & 25.9 & 25.8 & 25.7 & 25.6 & 25.8 & 26.4 & 26.7 & 27.0 & 27.0 \\
65425 &  0.4 & 0.75 & 215.132017 & 52.987357 & 28.0 & 28.0 & \dots & 27.8 & \dots & 27.6 & 28.1 & 27.4 & 27.8 & 28.8 & 28.6 & 28.8 & 28.9 \\
65435 &  0.4 & 0.75 & 215.110240 & 52.971622 & 24.8 & 22.6 & \dots & 21.9 & \dots & 21.8 & 21.5 & 21.4 & 21.6 & 22.2 & 22.5 & 22.6 & 22.7 \\
65465 &  2.8 & 1.00 & 215.057774 & 52.934783 & 28.1 & 27.6 & \dots & 27.4 & 27.0 & 27.0 & 27.5 & 27.1 & 26.3 & 25.7 & 26.0 & 25.9 & 25.5 \\
65491 &  2.8 & 1.00 & 215.114953 & 52.975189 & 24.5 & 23.7 & \dots & 23.2 & \dots & 23.1 & 23.2 & 23.0 & 23.2 & 23.7 & 24.2 & 24.4 & 24.5 \\
65525 &  0.4 & 0.75 & 215.126527 & 52.982928 & 29.0 & 29.5 & \dots & 28.6 & \dots & 29.9 & 28.9 & 28.9 & 27.6 & 27.8 & 28.7 & 27.7 & 28.3 \\
\hline
    \end{tabular}
    \caption{-- \textit{continued}}
\end{table*}

\setcounter{table}{0}

\begin{table*}
    \centering
    \begin{tabular}{l l l l l l l l l l l l l l l l l l}
ID & T & P(T) & RA & DEC & $m_{606}$ & $m_{814}$ & $m_{105}$ & $m_{125}$ & $m_{140}$ & $m_{160}$ & $m_{115}$ & $m_{150}$ & $m_{200}$ & $m_{277}$ & $m_{356}$ & $m_{410}$ & $m_{444}$ \\     
\hline
\hline
78102 &  0.4 & 0.75 & 215.050415 & 52.918988 & 28.2 & 27.7 & \dots & 27.4 & 26.6 & 27.3 & 26.9 & 26.8 & 27.0 & 27.6 & 27.9 & 28.1 & 28.2 \\
78167 &  0.4 & 0.75 & 214.989168 & 52.875528 & 24.4 & 22.1 & \dots & 21.2 & 21.1 & 21.1 & 21.1 & 21.0 & 21.2 & 21.7 & 21.8 & 21.9 & 22.1 \\
78273 &  0.4 & 0.75 & 214.982784 & 52.869861 & 28.1 & 28.0 & \dots & 28.0 & 27.6 & 27.6 & 27.6 & 27.6 & 27.5 & 27.7 & 28.1 & 28.4 & 28.5 \\
78700 &  0.4 & 0.75 & 215.023760 & 52.896850 & 28.9 & 27.8 & \dots & 27.1 & 26.8 & 27.1 & 27.1 & 26.9 & 27.2 & 27.6 & 28.1 & 28.1 & 28.4 \\
79318 &  0.4 & 0.75 & 215.038217 & 52.901973 & 26.0 & 22.2 & \dots & 20.6 & 20.6 & 20.5 & 20.6 & 20.5 & 20.5 & 21.0 & 21.1 & 21.1 & 21.3 \\
79826 &  0.4 & 0.75 & 215.039632 & 52.900466 & 25.9 & 23.8 & \dots & 23.1 & 23.1 & 23.0 & 23.0 & 22.8 & 23.0 & 23.5 & 23.8 & 24.0 & 24.1 \\
79850 &  0.4 & 0.75 & 214.984817 & 52.861576 & 27.5 & 27.5 & \dots & 27.4 & 26.5 & 26.4 & 27.1 & 26.4 & 26.9 & 27.4 & 27.3 & 27.7 & 27.5 \\
\hline
    \end{tabular}
    \caption{-- \textit{continued}}
\end{table*}

\setcounter{table}{0}

\begin{table*}
    \centering
    \begin{tabular}{l l l l l l l l l l l l l l l l l l}
ID & T & P(T) & RA & DEC & $m_{606}$ & $m_{814}$ & $m_{105}$ & $m_{125}$ & $m_{140}$ & $m_{160}$ & $m_{115}$ & $m_{150}$ & $m_{200}$ & $m_{277}$ & $m_{356}$ & $m_{410}$ & $m_{444}$ \\     
\hline
\hline
88543 &  0.4 & 0.75 & 214.937227 & 52.838216 & 23.9 & 23.4 & \dots & 23.8 & 24.1 & 24.1 & 23.5 & 23.9 & 24.4 & 25.0 & 25.6 & 26.0 & 26.0 \\
88742 &  0.4 & 0.75 & 214.914004 & 52.819424 & 25.5 & 24.4 & \dots & 24.1 & 24.1 & 24.0 & 24.0 & 23.9 & 24.1 & 24.7 & 25.0 & 25.2 & 25.4 \\
88919 &  0.4 & 0.75 & 214.881257 & 52.795377 & 25.1 & 24.3 & \dots & 24.0 & 24.0 & 23.9 & 24.0 & 23.8 & 24.0 & 24.5 & 24.9 & 25.1 & 25.2 \\
89143 &  0.4 & 0.75 & 214.917195 & 52.819380 & \dots & 29.1 & \dots & 28.4 & 27.1 & 27.3 & 27.3 & 27.5 & 28.0 & 28.2 & 28.2 & 28.7 & 28.6 \\
89305 &  0.4 & 0.75 & 214.960611 & 52.849277 & 28.6 & 27.7 & \dots & 26.3 & 26.0 & 26.4 & 24.3 & 24.5 & 24.9 & 25.4 & 26.0 & 26.3 & 26.5 \\
89885 &  0.4 & 0.75 & 214.950088 & 52.838719 & 26.0 & 25.5 & \dots & 25.2 & 25.3 & 25.2 & 25.3 & 25.2 & 25.5 & 25.9 & 26.5 & 26.7 & 26.9 \\
90078 &  0.4 & 0.75 & 214.949353 & 52.836889 & 26.0 & 24.5 & \dots & 23.9 & 23.9 & 23.8 & 23.8 & 23.7 & 23.9 & 24.3 & 24.6 & 24.8 & 25.0 \\
\hline
    \end{tabular}
    \caption{-- \textit{continued}}
\end{table*}

\setcounter{table}{0}

\begin{table*}
    \centering
    \begin{tabular}{l l l l l l l l l l l l l l l l l l}
ID & T & P(T) & RA & DEC & $m_{606}$ & $m_{814}$ & $m_{105}$ & $m_{125}$ & $m_{140}$ & $m_{160}$ & $m_{115}$ & $m_{150}$ & $m_{200}$ & $m_{277}$ & $m_{356}$ & $m_{410}$ & $m_{444}$ \\     
\hline
\hline
100326 &  2.8 & 1.00 & 214.856122 & 52.764319 & 27.1 & 26.0 & \dots & 25.5 & \dots & 25.6 & 25.5 & 25.4 & 25.7 & 26.1 & 26.4 & 26.7 & 26.7 \\
100445 &  0.8 & 0.75 & 214.852843 & 52.760909 & 26.4 & 23.6 & 30.4 & 22.4 & \dots & 22.3 & 21.9 & 21.8 & 22.0 & 22.5 & 22.7 & 22.9 & 23.0 \\
100731 &  0.4 & 0.50 & 214.835596 & 52.747057 & 28.1 & 28.4 & 30.8 & 28.7 & 28.0 & 27.5 & 27.9 & 27.4 & 27.9 & 28.3 & 28.2 & 28.3 & 28.3 \\
100756 &  0.4 & 0.75 & 214.868371 & 52.770407 & 28.8 & 29.3 & \dots & 28.0 & \dots & 29.0 & 29.1 & 28.7 & 27.7 & 28.3 & 28.8 & 28.3 & 29.0 \\
101332 &  2.8 & 1.00 & 214.860008 & 52.760677 & 21.8 & 21.7 & \dots & 21.9 & \dots & 21.9 & 21.7 & 21.8 & 22.2 & 22.8 & 23.3 & 23.6 & 23.7 \\
101361 &  0.4 & 0.75 & 214.867753 & 52.765822 & 21.1 & 20.7 & \dots & 20.7 & \dots & 20.7 & 20.6 & 20.7 & 20.9 & 21.5 & 22.0 & 22.2 & 22.3 \\
101624 &  0.4 & 0.75 & 214.884175 & 52.775749 & 28.3 & 26.0 & \dots & 25.2 & 25.1 & 25.2 & 24.9 & 24.8 & 25.0 & 25.4 & 25.7 & 25.8 & 25.9 \\
101633 &  0.4 & 0.75 & 214.876427 & 52.770284 & 21.0 & 20.9 & \dots & 21.0 & \dots & 21.1 & 20.9 & 21.3 & 21.5 & 21.9 & 22.5 & 22.7 & 22.8 \\
101761 &  0.4 & 0.75 & 214.888823 & 52.777875 & 24.0 & 23.1 & \dots & \dots & \dots & \dots & 22.8 & 22.7 & 22.9 & 23.4 & 23.8 & 24.0 & 24.1 \\
\hline
    \end{tabular}
    \caption{The CEERS unresolved sources identified in Figure \ref{f:ceers:mAB-r50} with kNN type $T$ and probability $P(T)$, RA and DEC position and the HST and JWST/NIRCam AB apparent magnitudes.}
    \label{t:ceers:stars}
\end{table*}

%%%%%%%%%%%%%%%%%%%%%%%%%%%%%%%%%%%%%%%%%%%%%%%%%%

% Don't change these lines
\bsp	% typesetting comment
\label{lastpage}
\end{document}